\documentclass[aps,twocolumn,showpacs,eqsecnum,superscriptaddress]{revtex4}

\usepackage{latexsym}
\usepackage{amssymb}
\usepackage{amsmath}
\usepackage{mathbbol}
\usepackage{epsfig}

\begin{document}

%%%%%%%%%%%%%%%%%
%%%   TITLE   %%%
%%%%%%%%%%%%%%%%%

\title{Generalized harmonic spatial coordinates and hyperbolic shift
conditions}

%%%%%%%%%%%%%%%%%%
%%%   AUTHORS  %%%
%%%%%%%%%%%%%%%%%%

\author{Miguel Alcubierre}
\affiliation{Instituto de Ciencias Nucleares, Universidad Nacional
Aut{\'o}noma de M{\'e}xico, A.P. 70-543, M{\'e}xico D.F. 04510, M{\'e}xico}

\author{Alejandro Corichi}
\affiliation{Instituto de Ciencias Nucleares, Universidad Nacional
Aut{\'o}noma de M{\'e}xico, A.P. 70-543, M{\'e}xico D.F. 04510, M{\'e}xico}

\author{Jos\'e A. Gonz\'alez}
\affiliation{Theoretical Physics Institute, University of Jena, 
Max-Wien-Platz 1, 07743, Jena, Germany}

\author{Dar\'\i o N\'u\~nez}
\affiliation{Instituto de Ciencias Nucleares, Universidad Nacional
Aut{\'o}noma de M{\'e}xico, A.P. 70-543, M{\'e}xico D.F. 04510, M{\'e}xico}

\author{Bernd Reimann}
\affiliation{Instituto de Ciencias Nucleares, Universidad Nacional
Aut{\'o}noma de M{\'e}xico, A.P. 70-543, M{\'e}xico D.F. 04510, M{\'e}xico}
\affiliation{Max Planck Institut f\"ur Gravitationsphysik,
Albert Einstein Institut, Am M\"uhlenberg 1, 14476 Golm, Germany}

\author{Marcelo Salgado}
\affiliation{Instituto de Ciencias Nucleares, Universidad Nacional
Aut{\'o}noma de M{\'e}xico, A.P. 70-543, M{\'e}xico D.F. 04510, M{\'e}xico}

%%%%%%%%%%%%%%%%
%%%   DATE   %%%
%%%%%%%%%%%%%%%%

\date{October 24, 2005.}

%%%%%%%%%%%%%%%%%%%%
%%%   ABSTRACT   %%%
%%%%%%%%%%%%%%%%%%%%

\begin{abstract}
We propose a generalization of the condition for harmonic spatial
coordinates analogous to the generalization of the harmonic time
slices introduced by Bona {\em et al.}, and closely related to dynamic
shift conditions recently proposed by Lindblom and Scheel, and Bona
and Palenzuela.  These generalized harmonic spatial coordinates imply
a condition for the shift vector that has the form of an evolution
equation for the shift components.  We find that in order to decouple
the slicing condition from the evolution equation for the shift it is
necessary to use a rescaled shift vector.  The initial form of the
generalized harmonic shift condition is not spatially covariant, but 
we propose a simple way to make it fully covariant so that it
can be used in coordinate systems other than Cartesian. We also
analyze the effect of the shift condition proposed here on the
hyperbolicity of the evolution equations of general relativity in 1+1
dimensions and 3+1 spherical symmetry, and study the possible
development of blow-ups.  Finally, we perform a series of numerical
experiments to illustrate the behavior of this shift condition.
\end{abstract}

%%%%%%%%%%%%%%%%
%%%   PACS   %%%
%%%%%%%%%%%%%%%%

\pacs{
04.20.Ex, % initial value problem
04.25.Dm, % numerical relativity
95.30.Sf  % relativity and gravitation
\qquad Preprint number: AEI-2005-108
}

%%%%%%%%%%%%%%%%%%%%%
%%%   MAKETITLE   %%%
%%%%%%%%%%%%%%%%%%%%%

\maketitle

%%%%%%%%%%%%%%%%%%%%%%%%
%%%   INTRODUCTION   %%%
%%%%%%%%%%%%%%%%%%%%%%%%

\section{Introduction}
\label{sec:introduction}

When one studies the time evolution of the gravitational field in
general relativity, a good choice of coordinates (a ``gauge'' choice)
can make the difference between finding a well behaved solution for a
large portion of the spacetime, or running into a coordinate (or
physical) singularity in a finite coordinate time, which would not
allow a numerical evolution to continue any further.  In the 3+1
formulation, the choice of the time coordinate is related with the
lapse function, while the choice of the spatial coordinates is related
to the shift vector.  Many different ways to choose the lapse and the
shift have been proposed and used in numerical simulations in the past
(see for example the pioneering papers of Smarr and
York~\cite{Smarr78a,Smarr78b}).  Some gauge choices involve solving
elliptic equations, while others involve solving evolution type
equations, which may or may not be hyperbolic in character.  Recently,
hyperbolic coordinate conditions have become a focus of attention, as
they in principle allow one to write the full set of dynamical
equations as a well-posed system~\cite{Sarbach:2002gr,
Alcubierre02a,Alcubierre02b,Alcubierre03b,Alcubierre03c,Lindblom:2003ad},
while at the same time being both easier to implement and considerably
less computationally expensive than elliptic conditions.

The classic example of hyperbolic coordinate conditions are the
so-called harmonic coordinates, which are defined by asking for the
wave operator acting on the coordinate functions ${x^\mu}$ to vanish.
Harmonic coordinate conditions have the important property of allowing
the Einstein field equations to be written as a series of wave
equations (with non-linear source terms) for the metric coefficients
$g_{\mu \nu}$.  Because of this, these conditions were used to prove
the first theorems on the existence of solutions to the Einstein
equations~\cite{Choquet52}.  This property of transforming the
Einstein equations into wave equations could in principle also be seen
as an important advantage in the numerical integration of these
equations.  Still, with few exceptions~(see for
example~\cite{Szilagyi02b,Garfinkle02a,Pretorius:2004jg}), full
harmonic coordinates have traditionally not been used in numerical
relativity, though harmonic time slices have been advocated and used
in some cases~\cite{Bona88,Bona89,Bona92,Abrahams95a}.  The reason for
this is two-fold: In the first place, harmonic coordinates are rather
restrictive, and formulations of the Einstein equations for numerical
relativity are usually written in a way that allows the gauge freedom
to remain explicit so it can be used to control certain aspects of the
evolution (avoid singularities, enforce symmetries, reduce shear,
etc.).  Also, in the particular case of a harmonic time coordinate, it
has been shown that the space-like foliation avoids focusing
singularities only marginally, and is therefore not a good choice in
many cases~\cite{Bona88,Geyer95,Bona97a,Alcubierre02b}.  Of course, it
can be argued that any coordinate choice is harmonic if one does not
ask for the wave operator acting on the coordinate functions to be
zero, but instead to be equal to a known function of spacetime (a
``gauge source function'').  This is certainly true, but of little use
in real life numerical simulations where there is no way to know {\em
a priori} what is a convenient choice for these gauge source functions
(but see~\cite{Pretorius:2004jg} for some suggestions that seem to
work well in practice).

Nevertheless, the fact that the use of harmonic coordinates allows the
field equations to be written in strongly hyperbolic form makes one
immediately ask if there might be simple generalizations of the
harmonic conditions that will still allow the field equations to be
written in strongly hyperbolic form, while at the same time retaining a 
useful degree of gauge freedom.  That this is indeed the case was first 
shown for the particular case of a harmonic time coordinate by Bona 
{\em et al.} in~\cite{Bona94b}, where a strongly hyperbolic reformulation 
of the Einstein evolution equations was constructed using a generalized
harmonic slicing condition which is usually referred to as the 
Bona-Masso slicing condition.  It includes as particular cases several 
choices that had been used in numerical simulations from the early 90's 
with good results, such as for example the ``1+log'' 
slicing~\cite{Bernstein93a,Anninos94c}.  In fact, the Bona-Masso slicing 
condition was motivated precisely to include such empirically tested 
conditions in a strongly hyperbolic formulation of the Einstein equations.

In this paper we want to follow a similar approach and propose a
generalization of the harmonic spatial coordinate condition.  We will
show how this allows us to obtain a hyperbolic shift condition that is
very closely related to conditions already proposed in the literature,
most notably the shift conditions recently introduced by Lindblom and
Scheel~\cite{Lindblom:2003ad}, and by Bona and
Palenzuela~\cite{Bona:2004yp} (in fact, under some specific
circumstances, one finds that the shift condition proposed here
becomes a particular case of those of Refs.~\cite{Lindblom:2003ad}
and~\cite{Bona:2004yp}).

This paper is organized as follows.  In Sec.~\ref{sec:harmonic} we
discuss the standard harmonic coordinates and write them as evolution
equations for the lapse and shift.  We also introduce a rescaled shift
vector that allows one to decouple the lapse and shift
equations. Section~\ref{sec:genharmonic} generalizes the condition for
spatial harmonic coordinates, and Sec.~\ref{sec:curvilinear} discusses
the interpretation of this condition in curvilinear coordinate
systems.  In Sec.~\ref{sec:hyperbolicity} we describe the concept of
hyperbolicity and the source criteria for avoiding blow-ups.
Section~\ref{sec:1p1} studies the generalized harmonic shift condition
in the case of 1+1 dimensions, analyzing its hyperbolicity properties,
the possible appearance of blow-ups (gauge shocks), and also the
behavior of this shift condition in numerical simulations.  In
Sec.~\ref{sec:sphsymm} we repeat the same type of analysis for
spherical symmetry and also present results from numerical
simulations.  We conclude in Sec.~\ref{sec:discussion}.  Finally,
Appendix~\ref{sec:appendix_modified} shows a formal derivation of the 
generalized harmonic lapse and shift conditions, and 
Appendix~\ref{sec:appendix_Christoffel} gives general expressions for the 
4-Christoffel symbols in terms of 3+1 quantities.

%%%%%%%%%%%%%%%%%%%%%%%%%%%%%%%%
%%%   HARMONIC COORDINATES   %%%
%%%%%%%%%%%%%%%%%%%%%%%%%%%%%%%%

\section{Harmonic coordinates}
\label{sec:harmonic}

Let us consider four scalar coordinate functions $\phi^\alpha$ defined
on a given background spacetime.  The condition for these coordinates
to be harmonic is simply
\begin{equation}
\Box \phi^\alpha := g^{\mu\nu} \,\nabla_\mu \nabla_\nu \, \phi^\alpha = 0 \; ,
\label{eq:harmonic1}
\end{equation}
with $g_{\mu\nu}$ the spacetime metric tensor.

Let us further assume that $\phi^0$ is such that its level surfaces
are space-like.  In that case, $\phi^0$ can be identified with a
global time function.  If we define the lapse function $\alpha$ as the
interval of proper time when going from the hypersurface $\phi^0=t$ to
the hypersurface $\phi^0=t+dt$ along the normal direction, then it is
easy to show that $\alpha$ will be given in terms of $\phi^0$ as
\begin{equation}
\alpha = \left( - \nabla \phi^0 \cdot \nabla \phi^0 \right)^{-1/2} \; .
\label{eq:alphadef}
\end{equation}

The definition of the shift vector is somewhat more involved.  We
start by defining three scalar functions $\beta^a$ such that when we
move from a given level surface of $\phi^0$ to the next following the
normal direction, the change in the spatial coordinate functions
$\phi^a$ is given by
\begin{equation}
\phi^a_{t+dt} = \phi^a_t - \beta^a d\phi^0 \; , 
\end{equation}
from which one can easily find
\begin{equation}
\beta^a = - \alpha \left( \vec{n} \cdot \nabla \phi^a \right) \; ,
\label{eq:betadef1}
\end{equation}
with $\vec{n}$ the unit normal vector to the hypersurface \mbox{$\phi^0=t$},
\begin{equation}
\vec{n} = - \alpha \nabla \phi^0 \; ,
\label{eq:normal}
\end{equation}
and where the minus sign is there to guarantee that $\vec{n}$ is future
pointing. Thus defined, the $\beta^a$ are scalars, but we can
use them to define a vector $\vec{\beta}$ by asking for its components
in the coordinate system $\{\phi^\alpha\}$ to be given by
$(0,\beta^a)$.  The vector constructed in this way is clearly
orthogonal to $\vec{n}$.  In an arbitrary coordinate system
$\{x^\mu\}$, the shift components will then be given by
\begin{equation}
\beta^\mu = - \alpha \left( \vec{n} \cdot \nabla \phi^a \right)
\frac{\partial x^\mu}{\partial \phi^a} \; .
\label{eq:betadef2}
\end{equation}

Notice that with this definition, the shift vector is proportional
to the lapse function, so that a simple rescaling of $\phi^0$ changes
the shift. This suggests that it is perhaps more natural to define a
rescaled shift vector $\vec{\sigma}$ in the following way
\begin{equation}
\sigma^\mu := \frac{\beta^\mu}{\alpha} =
- \left( \vec{n} \cdot \nabla \phi^a \right)
\frac{\partial x^\mu}{\partial \phi^a} \; .
\label{eq:sigmadef}
\end{equation}
We will see below that this rescaled shift vector will be important
when expressing the harmonic condition in 3+1 language.

The harmonic coordinate conditions can be simplified by expanding them
in the coordinate system $\{x^\alpha=\phi^\alpha\}$, in which case
they reduce to
\begin{equation}
\Gamma^\alpha := g^{\mu \nu} \Gamma^\alpha_{\mu \nu} = 0 \; ,
\label{eq:harmonic2}
\end{equation}
where $\Gamma^\alpha_{\mu \nu}$ are the Christoffel symbols associated with
the 4-metric $g_{\mu \nu}$.  If we now relate the coordinates
\mbox{$\{x^\alpha=\phi^\alpha\}$} to the standard 3+1 coordinates,
then these four equations can be shown to become (see
Appendix~\ref{sec:appendix_modified} and \ref{sec:appendix_Christoffel})
\begin{eqnarray}
\partial_t \alpha  &=& \beta^a \partial_a \alpha - \alpha^2 K \; ,
\label{eq:alphadothar} \\
\partial_t \beta^i &=& \beta^a \partial_a \beta^i - \alpha \partial^i
\alpha + \alpha^2 \; {}^{(3)}\Gamma^i \nonumber \\
&& + \frac{\beta^i}{\alpha}
\left( \partial_t \alpha - \beta^a \partial_a \alpha
+ \alpha^2 K \right) \; .
\label{eq:betadothar}
\end{eqnarray}
Here $K$ is the trace of the extrinsic curvature, and
${}^{(3)}\Gamma^i$ is defined in terms of the three-dimensional
Christoffel symbols ${}^{(3)}\Gamma^i_{jk}$, the spatial metric
$\gamma_{ij}$ and its determinant \mbox{$\gamma:={\rm
det}\:\gamma_{ij}$} by \mbox{${}^{(3)}\Gamma^i := \gamma^{jk}
{}^{(3)}\Gamma^i_{jk} = - \partial_j\left(\sqrt{\gamma} \:
\gamma^{ij}\right) / \sqrt{\gamma}$}.  Notice that in
equation~(\ref{eq:betadothar}) we have an explicit dependency on the
time derivative of the lapse function.  This dependency is usually not
written down, as the whole last term of the second equation vanishes
if the first equation is assumed to hold, but we prefer to leave the
dependency explicit (see for example~\cite{York79,Garfinkle02a};
incidentally, equation~(\ref{eq:betadothar}) fixes a sign error
in~\cite{York79}, and includes a term missing in~\cite{Garfinkle02a}).

The fact that the evolution equation for the shift depends on the time
derivative of the lapse is inconvenient if one wants to use harmonic
spatial coordinates with a different slicing condition, say maximal
slicing.  It is also an indication that the shift itself might not be
the most convenient function to evolve.  Remarkably, it turns out that
if we rewrite the evolution equation for the shift in terms of the
rescaled shift $\sigma^i=\beta^i/\alpha$ introduced above, then the
spatial harmonic condition decouples completely from the evolution of
the lapse.  We find
\begin{equation}
\partial_t \sigma^i = \alpha \sigma^a \partial_a \sigma^i
- \partial^i \alpha + \alpha \left( \sigma^i K
+ {}^{(3)}\Gamma^i \right) \; .
\label{eq:sigmadothar}
\end{equation}
Therefore, if one works with $\sigma^i$ instead of $\beta^i$, one can
use harmonic spatial coordinates with an arbitrary slicing condition
in a straightforward way.

A final comment about equations~(\ref{eq:alphadothar})
and~(\ref{eq:betadothar}) is in order.
Equation~(\ref{eq:alphadothar}) is clearly a scalar equation as seen
in the spatial hypersurfaces.  Equation~(\ref{eq:betadothar}), on the
other hand, is not 3-covariant, {\em i.e.} starting from exactly the
same 3-geometry but in different coordinates, it will produce a
different evolution for the shift vector.  This might seem surprising
since this equation is just the 3+1 version of the condition for
spatial harmonic coordinates which is 4-covariant.  However, there is
no real contradiction, since changing the coordinates on the spatial
hypersurfaces means changing the scalar functions $\phi^i$ themselves,
so it should not be surprising that we get a different shift.  We will
come back to this point in Sec.~\ref{sec:curvilinear}, where we
will propose a way to make the shift evolution equation fully
3-covariant.

%%%%%%%%%%%%%%%%%%%%%%%%%%%%%%%%
%%%   GENERALIZED HARMONIC   %%%
%%%%%%%%%%%%%%%%%%%%%%%%%%%%%%%%

\section{Generalized harmonic coordinates}
\label{sec:genharmonic}

In~\cite{Bona94b}, Bona {\em et al.} generalize the harmonic slicing
condition~(\ref{eq:alphadothar}) in the following way
\begin{equation}
\partial_t \alpha - \beta^a \partial_a \alpha = - \alpha^2 f(\alpha) K
\; ,
\label{eq:alphaBM}
\end{equation}
with $f(\alpha)$ a positive but otherwise arbitrary function of the
lapse.  This slicing condition was originally motivated by the
Bona-Masso hyperbolic reformulation of the Einstein
equations~\cite{Bona89,Bona92,Bona93,Bona94b,Bona97a}, but it can in
fact be used with any form of the 3+1 evolution equations.  As
discussed in~\cite{Alcubierre02b}, the Bona-Masso slicing condition
above can be shown to avoid both focusing singularities~\cite{Bona97a}
and gauge shocks~\cite{Alcubierre97a} for particular choices of $f$.
Reference~\cite{Alcubierre02b} also shows that
condition~(\ref{eq:alphaBM}) can be written in 4-covariant form in
terms of a global time function $\phi^0$ as
\begin{equation}
\left( g^{\mu\nu} - a_f \, n^\mu n^\nu \right) \nabla_\mu \nabla_\nu \phi^0
= 0 \; ,
\label{eq:GHtime}
\end{equation}
with $a_f := 1/f(\alpha)-1$ and $n^\mu$ the unit normal vector to the
spatial hypersurfaces defined in~(\ref{eq:normal}).  Here we will
introduce an analogous generalization of the spatial harmonic
coordinates $\{\phi^l\}$. That is, we propose the following spatial
gauge condition
\begin{equation}
\left( g^{\mu\nu} - a_h \, n^\mu n^\nu \right) \nabla_\mu \nabla_\nu \phi^l
= 0 \; ,
\label{eq:GHspatial}
\end{equation}
where $n^\mu$ is still the unit normal to the spatial hypersurfaces,
but now $a_h := 1/h-1$, with $h(\alpha,\beta^i)$ a scalar function
that can in principle depend on both the lapse and shift (we will see
below that the shift dependence is in fact not convenient).  In the
coordinate system $\{x^\mu = \phi^\mu\}$,
condition~(\ref{eq:GHspatial}) becomes
\begin{equation}
\left( g^{\mu\nu} - a_h \, n^\mu n^\nu \right) \Gamma_{\mu\nu}^l
= 0 \; .
\label{eq:GHspatial2}
\end{equation}
Expressing the 4-metric and normal vector in terms of 3+1 variables, 
the last equation becomes
\begin{equation}
\Gamma_{00}^l - 2 \beta^m \Gamma_{m0}^l + \beta^m \beta^n \Gamma_{mn}^l
= \alpha^2 h \; \gamma^{mn} \Gamma^l_{mn} \; .
\label{eq:GHspatial3}
\end{equation}
Notice that on the right hand side of this equation appears the
contraction $\gamma^{mn} \Gamma^l_{mn}$ which should not be confused
with \mbox{$\Gamma^l:= g^{\mu\nu} \Gamma^l_{\mu\nu}$}.  Inserting now
the expressions for the $\Gamma^l_{mn}$ in terms of 3+1 quantities we
obtain (see also Appendix~\ref{sec:appendix_Christoffel})
\begin{eqnarray}
\partial_t \beta^l &=& \beta^m \partial_m \beta^l - \alpha \partial^l \alpha
+ \frac{\beta^l}{\alpha} \left( \partial_t \alpha
- \beta^m \partial_m \alpha \right) \hspace{5mm} \nonumber \\
&+& \alpha^2 h \left( \frac{\beta^l}{\alpha} K
+ {}^{(3)}\Gamma^l \right) \; .
\label{eq:betadotGH}
\end{eqnarray}
This is to be compared with equation~(\ref{eq:betadothar}) of the
previous section.  Notice that again we find that the evolution
equation for the shift is coupled to that of the lapse. In the same
way as before, we can decouple the shift evolution equation by writing
it in terms of the rescaled shift $\sigma^i=\beta^i/\alpha$. We find
\begin{equation}
\partial_t \sigma^l = \alpha \sigma^m \partial_m \sigma^l
- \partial^l \alpha + \alpha h \left( \sigma^l K
+ {}^{(3)}\Gamma^l \right) \; ,
\label{eq:sigmadotGH}
\end{equation}
which is to be compared with~(\ref{eq:sigmadothar}).  This is the
final form of the condition for generalized harmonic spatial
coordinates, and we will refer to this condition simply as the
``generalized harmonic shift'' (but see Sec.~\ref{sec:curvilinear}
below where the condition is somewhat modified to make it fully
3-covariant).

At this point it is important to discuss the relation that the shift
condition~(\ref{eq:betadotGH}) has with the conditions recently
proposed by Lindblom and Scheel~\cite{Lindblom:2003ad}, and by Bona
and Palenzuela~\cite{Bona:2004yp}.  It is not difficult to see that by
choosing the free parameters in these references appropriately, one
can in fact recover condition~(\ref{eq:betadotGH}), but only {\em
provided}\/ one also takes the lapse to evolve via the Bona-Masso
slicing condition~(\ref{eq:alphaBM}) and takes $f=h$.  If, on the
other hand, one uses a different slicing condition (say maximal
slicing), or uses the Bona-Masso slicing condition with $f\neq h$,
then this is no longer the case and the shift condition proposed here
will differ from those of Refs.~\cite{Lindblom:2003ad}
and~\cite{Bona:2004yp}.  This is a crucial point, and shows the
importance of rescaling the shift in order to decouple its evolution
equation from the time derivative of the lapse.

In the following sections we will study this shift condition.  We will
first discuss the issue of the interpretation of the generalized
harmonic shift condition for curvilinear coordinates in
Sec.~\ref{sec:curvilinear}. Later, in Sec.~\ref{sec:hyperbolicity} we
will introduce the concept of hyperbolicity, and a criteria for
avoiding blow-ups in the solutions of strongly hyperbolic systems of
equations.  Finally, in Sections~\ref{sec:1p1} and~\ref{sec:sphsymm}
we will consider the special cases of 1+1 dimensional relativity and
spherical symmetry.  In each case we will analyze the hyperbolicity
properties of the full system of equations {\em including}\/ the
generalized harmonic shift condition, study the possible development
of blow-ups, and present a series of numerical examples.

%%%%%%%%%%%%%%%%%%%%%%%%%%%%%%%%%%%
%%%   CURVILINEAR COORDINATES   %%%
%%%%%%%%%%%%%%%%%%%%%%%%%%%%%%%%%%%

\section{Curvilinear versus Cartesian coordinates}
\label{sec:curvilinear}

We have already mentioned that the harmonic shift
condition~(\ref{eq:sigmadothar}), and its
generalization~(\ref{eq:sigmadotGH}), are in fact not covariant with
respect to changes in the spatial coordinates.  That is, starting from
exactly the same 3-geometry but with different spatial coordinates we
will get a different evolution of the shift vector.  In particular,
for curvilinear systems of coordinates one could find that even
starting from a flat slice of Minkowski spacetime we would still have
non-trivial shift evolution driven by the fact that the initial
$^{(3)}\Gamma^i$ do not vanish ({\em i.e.} the spatial curvilinear
coordinates are not 3-harmonic).  Worse still, in many cases it can
happen that the $^{(3)}\Gamma^i$ of flat space are not only non-zero
but are also singular, as is the case with spherical coordinates for
which $^{(3)}\Gamma^r$ is of order $1/r$. One may also find that in
physical systems that have a specific symmetry the shift evolution
will break the symmetry because of the properties of some of the
$^{(3)}\Gamma^i$.  An example of this are again spherical coordinates
for which one finds that $^{(3)}\Gamma^\theta \neq 0$, so
$\sigma^\theta$ will evolve away from zero even for a spherically
symmetric system.

The question then arises how to interpret the harmonic shift
condition in a general coordinate system, and in particular how to make
sure that we do not run into pathological situations like those
described above.  Our proposal for resolving this issue is to always
apply the generalized harmonic shift condition in a coordinate system
that is topologically Cartesian.  Of course, if one has a situation
that has a specific symmetry, one would like to work with a coordinate
system that is adapted to that symmetry.  We therefore need to
transform condition~(\ref{eq:sigmadotGH}) from Cartesian coordinates
to our curvilinear coordinates, but taking into account the fact that
the condition is not covariant.

Let us denote by $\{x^{\bar{a}}\}$ our reference topologically
Cartesian coordinates, and by $\{x^i\}$ the general curvilinear
coordinates.  If we assume that condition~(\ref{eq:sigmadotGH}) is
satisfied for the original coordinates $\{x^{\bar{a}}\}$ we will have
\begin{equation}
\partial_t \sigma^{\bar{a}} = \alpha \sigma^{\bar{b}}
\partial_{\bar{b}} \sigma^{\bar{a}}
- \partial^{\bar{a}} \alpha + \alpha h \left( \sigma^{\bar{a}} K
+ {}^{(3)}\Gamma^{\bar{a}} \right) \; .
\label{eq:sigmadot_cartesian}
\end{equation}
In order to transform this expression we will use the fact that with
respect to the 3-geometry $\sigma^i$ behaves like a vector, while
$\alpha$ and $K$ behave as scalars.  Remembering now that the
Christoffel symbols transform as
\begin{equation}
^{(3)}\Gamma^i_{jk} = \left( \partial_{\bar{a}} x^i \; \partial_j x^{\bar{b}}
\; \partial_k x^{\bar{c}} \right) {}^{(3)}\Gamma^{\bar{a}}_{\bar{b}\bar{c}}
+ F^i_{jk} \; ,
\end{equation}
with $F^i_{jk} := \partial_{\bar{a}} x^i \; \partial_j \partial_k
x^{\bar{a}}$, we find that in the curvilinear coordinate system
equation~(\ref{eq:sigmadot_cartesian}) becomes
\begin{eqnarray}
\partial_t \sigma^l &=& \alpha \sigma^m \partial_m \sigma^l + \alpha
\sigma^m \sigma^n F^l_{mn} - \partial^l \alpha \nonumber \\ &+& \alpha
h \left( \sigma^l K + {}^{(3)}\Gamma^l - \gamma^{mn} F^l_{mn} \right)
\, .
\end{eqnarray}
By rearranging some terms, the shift condition can finally be written
in the more convenient form
\begin{eqnarray}
\partial_t \sigma^l &=& \alpha \sigma^m \nabla_m \sigma^l - \nabla^l
\alpha + \alpha h \: \sigma^l K \nonumber \\ &+& \alpha \left( h
\gamma^{mn} - \sigma^m \sigma^n \right) \Delta^l_{mn} \; ,
\hspace{8mm}
\label{eq:sigmadot_curve}
\end{eqnarray}
with $\Delta^l_{mn} := {}^{(3)}\Gamma^l_{mn} - F^l_{mn}$.  The last
expression is in fact 3-covariant, as one can readily verify that the
$\Delta^l_{mn}$ transform as the components of a 3-tensor.  But the
price we have paid is that we have chosen a privileged coordinate
system to be used as a reference in order to define $F^l_{mn}$.  It is
clear that for the original coordinates $\{x^{\bar{a}}\}$ the
condition above reduces to what we had before since $F^l_{mn}$
vanishes.  We will consider the case of spherical coordinates in
Sec.~\ref{sec:sphsymm} below.

In practice, one can use the fact that for flat space in Cartesian
coordinates the Christoffel symbols vanish, which implies
\begin{equation}
F^l_{mn} = \left. ^{(3)}\Gamma^l_{mn} \right|_{\rm flat} \, ,
\end{equation}
so that
\begin{equation}
\Delta^l_{mn} = {}^{(3)}\Gamma^l_{mn}
- \left. {}^{(3)}\Gamma^l_{mn} \right|_{\rm flat}.
\end{equation}

%%%%%%%%%%%%%%%%%%%%%%%%%%%%%%%%%%%%
%%%   HYPERBOLICITY AND SHOCKS   %%%
%%%%%%%%%%%%%%%%%%%%%%%%%%%%%%%%%%%%

\section{Hyperbolicity and shocks}
\label{sec:hyperbolicity}

%%%%%%%%%%%%%%%%%%%%%%%%%
%%%   HYPERBOLICITY   %%%
%%%%%%%%%%%%%%%%%%%%%%%%%

\subsection{Hyperbolic systems}
\label{sec:hyperbolic}

The concept of hyperbolicity is of fundamental importance in the study
of the evolution equations associated with a Cauchy problem as the
initial value problem for strongly or symmetric hyperbolic systems can
be shown to be well-posed (though the well-posedness of strongly
hyperbolic systems requires that some additional smoothness conditions
are verified).  In the following we will concentrate on
one-dimensional systems, for which the distinction between strongly
and symmetric hyperbolic systems does not arise.

Following \cite{Reimann:2004wp}, we will consider quasi-linear systems
of evolution equations that can be split into two subsystems of the
form
\begin{eqnarray}
\partial_t u &=& {\bf M}(u) \; v \; , 
\label{eq:uPDE} \\
\partial_t v &+& {\bf A}(u) \; \partial_x v
= q_v(u,v) \; .
\label{eq:vPDEmatrix}
\end{eqnarray}
Here $u$ and $v$ are $n$ and $m$ dimensional vector-valued functions,
and ${\bf M}$ and ${\bf A}$ are $n \times n$ and $m \times m$
matrices, respectively.  In addition we demand that the $v$'s are
related to either time or space derivatives of the $u$'s.  This
implies that derivatives of the $u$'s can always be substituted for
$v$'s and hence treated as source terms.

The system of equations above will be hyperbolic if the matrix
$\bf{A}$ has $m$ real eigenvalues $\lambda_i$.  Furthermore, it will
be strongly hyperbolic if it has a complete set of eigenvectors
$\vec{\xi}_i$.  If we denote the matrix of column eigenvectors by
${\bf R} = ( \vec{\xi}_1 \cdots \vec{\xi}_m )$, then the matrix
$\bf{A}$ can be diagonalized as
\begin{equation}
{\bf R}^{-1} {\bf A R} = \begin{rm}{\textbf{diag}}\end{rm}
\left[ \lambda_1, \cdots, \lambda_m \right]
= {\bf \Lambda} \; .
\label{eq:diagonalize}
\end{equation}
For a strongly hyperbolic system we then define the eigenfields as
\begin{equation}
\label{eq:w}
w = {\bf R}^{-1} v \; .
\end{equation}

By analyzing the time evolution of the eigenfields, one can identify
mechanisms that lead to blow-ups in the solution, which
in~\cite{Alinhac} have been referred to as ``geometric blow-up''
(leading to ``gradient catastrophes''~\cite{John86}) and the
``ODE-mechanism'' (causing ``blow-ups within finite time'').
In~\cite{Reimann:2004wp} some of us presented blow-up avoiding
conditions for both these mechanisms, which we called ``indirect
linear degeneracy''~\cite{Alcubierre97a} and the ``source criteria''.
In that reference it was also shown, using numerical examples, that
the source criteria for avoiding blow-ups is generally the more
important of the two conditions.  Because of this, and also because of
the fact that the true relevance of indirect linear degeneracy is not
as yet completely clear, in this paper we will concentrate only on the
source criteria.

%%%%%%%%%%%%%%%%%%%%%%%%%%%
%%%   SOURCE CRITERIA   %%%
%%%%%%%%%%%%%%%%%%%%%%%%%%%

\subsection{Source criteria for avoiding blow-ups}
\label{subsec:sourcecriteria}

An evolution variable can become infinite at a given point by a
process of ``self-increase'' in the causal past of this point.  A
criteria to avoid such blow-ups for systems of partial differential
equations of the form (\ref{eq:uPDE}) - (\ref{eq:vPDEmatrix}) was
proposed by some of us in~\cite{Reimann:2004wp}: When diagonalizing
the evolution system for the $v$'s, making use of
(\ref{eq:diagonalize}) and (\ref{eq:w}), one finds
\begin{equation}
\label{eq:wPDE}
\partial_t w + {\bf \Lambda} \; \partial_x w
= q_w \; ,
\end{equation}
where
\begin{equation}
q_w := {\bf R}^{-1} q_v + \left[ \partial_t  {\bf R}^{-1}
+ {\bf \Lambda} \partial_x  {\bf R}^{-1} \right] v \; .
\label{eq:qw}
\end{equation}
This yields an evolution system where on the left-hand side of
(\ref{eq:wPDE}) the different eigenfields $w_i$ are
decoupled. However, in general the equations are still coupled through
the source terms $q_{w_i}$.  In particular, if the original sources
were quadratic in the $v$'s, one obtains
\begin{equation}
\frac{dw_i}{dt} = \partial_t w_i + \lambda_i \partial_x
w_i = \sum_{j,k=1}^m c_{ijk} w_j w_k + {\cal O}(w) \; ,
\end{equation}
where $d/dt := \partial_t + \lambda_i \partial_x$ denotes the
derivative along the corresponding characteristic.  As pointed out
in~\cite{Reimann:2004wp}, the $c_{iii} w_i^2$ component of the source
term can be expected to dominate and to cause blow-ups in the solution
within a finite time.  In order to avoid these blow-ups we therefore
demand that the coefficients $c_{iii}$ should vanish, and we refer to
this condition as the ``source criteria''.

It is not difficult to convince one-self that the source criteria is
in fact not a sufficient condition for avoiding blow-ups, as already
discussed in~\cite{Reimann:2004wp}.  However, one can still expect 
the source criteria to be a necessary condition for avoiding
blow-ups at least for small perturbations propagating with different
eigenspeeds, as mixed terms will be suppressed when pulses moving at
different speeds separate from each other, while the effect of the
term $c_{iii} w_i^2$ will remain as each pulse moves.  If, however,
some eigenfields $w_i$ and $w_j$ travel with identical or similar
eigenspeeds, then one should also expect important contributions
coming from the mixed terms $w_i w_j$.  We will show an example later
on where eliminating such mixed terms (in addition to the quadratic
terms) indeed leads to further improvements.

%%%%%%%%%%%%%%%%%%%%%%
%%%   1+1 SYSTEM   %%%
%%%%%%%%%%%%%%%%%%%%%%

\section{Einstein equations in 1+1 dimensions}
\label{sec:1p1}

We first consider standard general relativity in one spatial dimension
(and in vacuum). Since in this paper we are interested precisely in
studying a new shift condition, 1+1 dimensional relativity is an ideal
testing ground for the ``gauge dynamics'' which one can expect in the 
higher dimensional case.

In the following sections we will introduce the evolution equations
and gauge conditions, and consider the possible formation of blow-ups
associated with our gauge conditions.  We will also present numerical
simulations that show how the generalized harmonic shift condition
behaves in practice.

%%%%%%%%%%%%%%%%%%%%%%%%%
%%%   1+1 EQUATIONS   %%%
%%%%%%%%%%%%%%%%%%%%%%%%%

\subsection{Evolution equations}
\label{subsec:1p1equations}

We will start from the ``standard'' Arnowitt-Deser-Misner (ADM)
equations for one spatial dimension~\cite{Arnowitt62} as formulated 
in~\cite{York79}.  In this case the $u$ quantities consist of the lapse 
function $\alpha$, the rescaled shift $\sigma := \sigma^x$, and the spatial 
metric function $g := g_{xx}$.  The $v$ quantities, on the other hand, are 
given by the spatial derivatives of the $u$'s and, in addition, the unique
component of the extrinsic curvature.  That is,
\begin{equation}
u = \left( \alpha , \sigma, g \right) \; , \qquad
v = \left( D_\alpha , d_\sigma, D_g , \tilde{K} \right) \; , 
\end{equation}
with $\tilde{K} := \sqrt{g} \: {\rm tr} K = \sqrt{g} \: K_x^x$, and
where we have defined
\begin{equation}
D_\alpha := \partial_x \ln \alpha \; , \quad
D_g := \partial_x \ln g \; , \quad
d_\sigma := \partial_x \sigma \: .
\end{equation}
Notice first that we use a rescaled extrinsic curvature, as this makes
the evolution equations considerably simpler.  Also, we use
logarithmic spatial derivatives of $\alpha$ and $g$, but only the
ordinary spatial derivative of the rescaled shift $\sigma$, as the
shift is allowed to change sign.

For the evolution of the gauge variables we will use the Bona-Masso
slicing condition~(\ref{eq:alphaBM}) and the generalized harmonic
shift condition~(\ref{eq:sigmadotGH}). The equations for the $u$'s are
then
\begin{eqnarray}
\partial_t \alpha &=&  \alpha^2 \left[
\sigma D_\alpha - \frac{f \tilde{K}}{\sqrt{g}} \right] \; ,
\label{eq:alphadot_1p1} \\
\partial_t \sigma &=& \alpha \left[ \sigma d_\sigma
- \frac{D_\alpha}{g}
+ h \left( \frac{D_g}{2g} + \frac{\sigma \tilde{K}}{\sqrt{g}} \right) \right]
\; , \qquad
\label{eq:sigmadot_1p1} \\ 
\partial_t g &=& \alpha g \left[
\sigma D_g + 2 \left( d_\sigma + \sigma D_\alpha
- \frac{\tilde{K}}{\sqrt{g}} \right) \right] \; ,
\label{eq:gdot_1p1}
\end{eqnarray}
where $f = f(\alpha)$ and $h = h(\alpha,\sigma)$.  The evolution
equations for $\{D_\alpha,d_\sigma,D_g\}$ can be obtained directly
from the above equations, while the evolution equation for $\tilde{K}$
comes from the ADM equations and takes the following simple form
\begin{equation}
\partial_t \tilde{K} = \partial_x \left[ \alpha \left(
\sigma \tilde{K} - D_\alpha / \sqrt{g} \right) \right] \; .
\label{eq:Kgdot_1p1}
\end{equation}

The evolution equations for the $v$'s can then be written in full
conservative form \mbox{$\partial_t v + \partial_x \left( {\bf A} \: v
\right) = 0$}, with the characteristic matrix ${\bf A}$ given by
\begin{equation}
{\bf A} = \left( \begin{array}{cccc}
- \alpha \sigma & 0 & 0 & \alpha f / \sqrt{g} \\
\alpha/g & - \alpha \sigma & - \alpha h / 2g & - \alpha h \sigma / \sqrt{g} \\
- 2 \alpha \sigma & - 2 \alpha & - \alpha \sigma & 2 \alpha / \sqrt{g} \\
\alpha / \sqrt{g} & 0 & 0 & - \alpha \sigma
\end{array} \right) \; .
\label{eq:A_1p1}
\end{equation}
This matrix has the following eigenvalues
\begin{eqnarray}
\lambda^f_\pm &=& \alpha \left( \pm \sqrt{f/g} 
- \sigma \right) \: , 
\label{eq:lambdaf_1p1} \\
\lambda^h_\pm &=& \alpha \left( \pm \sqrt{h/g} 
- \sigma \right) \: ,
\label{eq:lambdah_1p1}
\end{eqnarray}
with corresponding eigenfunctions (the normalization is chosen for
convenience)
\begin{eqnarray}
w^f_\pm &=& \tilde{K} \pm D_\alpha / \sqrt{f} \: ,
\label{eq:wfpm_1p1} \\
w^h_\pm &=& - \left(1 \pm \sigma \sqrt{g h} \right) \tilde{K}
+ \sqrt{g} \: d_\sigma \mp \sqrt{h} D_g /2 \: , \hspace{10mm}
\label{eq:whpm_1p1}
\end{eqnarray}
which can be easily inverted to find
\begin{eqnarray}
D_\alpha &=& \frac{\sqrt{f}}{2} \left( w^f_+ - w^f_- \right) \; ,
\label{eq:Dalpha_eigen} \\
d_\sigma &=& \frac{1}{2 \sqrt{g}} 
\left( w^f_+ + w^f_- + w^h_+ + w^h_- \right) \; ,
\label{eq:Dsigma_eigen} \\
D_g &=& \frac{1}{\sqrt{h}} \left( w^h_- - w^h_+ \right)
- \sigma \sqrt{g} \left( w^f_+ + w^f_- \right) \; , \qquad 
\label{eq:Dg_eigen} \\
\tilde{K} &=& \frac{1}{2} \left( w^f_+ + w^f_- \right) \; .
\label{eq:Ktilde_eigen} 
\end{eqnarray}

The system is therefore strongly hyperbolic as long as $f>0$ and
$h>0$, with the lapse and shift eigenfields $w^f_\pm$ and $w^h_\pm$
propagating with the corresponding gauge speeds $\lambda^f_{\pm}$ and
$\lambda^h_\pm$.

%%%%%%%%%%%%%%%%%%%%%%%%
%%%   1+1 ANALYSIS   %%%
%%%%%%%%%%%%%%%%%%%%%%%%

\subsection{Gauge shock analysis}
\label{subsec:1p1analysis}

By analyzing quadratic source terms in the evolution equations of the 
eigenfields $w_i$, we now want to study the possible formation of blow-ups 
for the system of evolution equations of the previous section.  For the 
lapse and shift eigenfields we find
\begin{eqnarray}
\frac{d w^f_\pm}{dt} 
& = & c^{fff}_{\pm\pm\pm} w^{f \: 2}_\pm 
+ c^{ffh}_{\pm\pm\pm} w^f_\pm w^h_\pm \nonumber \\
& + & {\cal O} \left( w^f_\pm w^f_\mp , w^f_\pm w^h_\mp \right) \; , 
\label{eq:dwfdt_1p1} \\
\frac{d w^h_\pm}{dt} 
& = & c^{hhh}_{\pm\pm\pm} w^{h \: 2}_\pm 
+ c^{hhf}_{\pm\pm\pm} w^h_\pm w^f_\pm \nonumber \\
& + & {\cal O} \left( w^h_\pm w^h_\mp , w^h_\pm w^f_\mp \right) \; .
\label{eq:dwhdt_1p1}
\end{eqnarray}
In particular, we observe that in (\ref{eq:dwfdt_1p1}) no term
proportional to $w^{h \: 2}_\pm$ is present, and in the same way in
(\ref{eq:dwhdt_1p1}) there is no term proportional to $w^{f \:
2}_\pm$.  In order to apply the source criteria we need to calculate those 
terms quadratic in $w_i$ appearing in the sources of the evolution equation 
for $w_i$ itself.  It turns out that the $c_{iii}$ coefficients have the form
\begin{eqnarray}
c^{fff}_{\pm\pm\pm} &\propto& 
\left( 1 - f - \alpha f' / 2 \right) \; , \\
c^{hhh}_{\pm\pm\pm} &\propto& 
\partial h / \partial \sigma \; .
\end{eqnarray}
According to the source criteria these coefficients have to vanish in
order to avoid blow-ups.  The conditions on the gauge functions
$f(\alpha)$ and $h(\alpha,\sigma)$ are then
\begin{eqnarray}
1 - f - \alpha f' / 2 &=& 0
\label{eq:fcond_1p1} \; , \\
\partial h / \partial \sigma &=& 0
\label{eq:hcond_1p1} \; .
\end{eqnarray}
The condition~(\ref{eq:fcond_1p1}) for $f(\alpha)$ has been studied
many times
before~\cite{Alcubierre97a,Alcubierre97b,Alcubierre02b,Reimann:2004wp},
and its general solution is
\begin{equation}
f(\alpha) = 1 + {\rm const} / \alpha^2 \; .
\label{eq:shockavoidf}
\end{equation}

For $h(\alpha,\sigma)$, on the other hand, we obtain the condition
that $h$ can be an arbitrary function of $\alpha$, but may not
depend on $\sigma$, that is, $h = h(\alpha)$.

One now might wonder about the case where $h$ is equal (or very close
to) the function $f$.  In that case the eigenfields $w^f_\pm$ and
$w^h_\pm$ travel with the same (or similar) eigenspeeds, so mixed
terms of the type $w^f_\pm w^h_\pm$ in the sources can be expected to
contribute to a blow-up.  For this reason we have also calculated the
$c_{iij}$ coefficients associated to these terms.  Notice, however, that 
in general the coefficients of such mixed terms are not invariant under 
rescalings of the eigenfields of the form $\tilde{w}_i = \Omega_i 
(\alpha,\sigma,g) \: w_i$, so we have in fact done the calculation 
assuming an arbitrary rescaling.  We find
\begin{eqnarray}
 c^{ffh}_{\pm\pm\pm} &\propto& 
\left( 1 - \sqrt{\frac{h}{f}} \right) , \\
 c^{hhf}_{\pm\pm\pm} &\propto& 
\left\{ \left[ 2 \alpha \sqrt{f} \: 
\frac{\partial \Omega^h_\pm}{\partial \alpha} 
\pm \frac{2}{\sqrt{g}} \frac{\partial \Omega^h_\pm}{\partial \sigma} 
\right. \right. \nonumber \\
& \mp & \left. \left. 4 \sigma \sqrt{g} 
\left( \Omega^h_\pm + g \frac{\partial \Omega^h_\pm}{\partial g} \right) 
\right] \left( 1 - \sqrt{\frac{h}{f}} \right) \right. \nonumber \\
& + & \left. \Bigg[ \frac{\sqrt{f}}{2h} \left( 1 + \sqrt{\frac{h}{f}} \right)
\left( \alpha \frac{\partial h}{\partial \alpha} 
\pm \frac{1}{\sqrt{g f}} \frac{\partial h}{\partial \sigma} \right)
\right. \nonumber \\
& + & \left. \frac{1+3h}{\sqrt{h}} - \frac{3+h}{\sqrt{f}}
\Bigg] \Omega^h_\pm \right\} \; . \qquad
\end{eqnarray}

One can readily verify that these coefficients vanish for $f = h = 1 +
{\rm const} / \alpha^2$, independently of the rescaling
of the eigenfields.  This setting of $f$ and $h$ hence seems to be an
optimal choice for avoiding blow-ups.

%%%%%%%%%%%%%%%%%%%%%%%%%%%%%%%%%%
%%%   1+1 NUMERICAL EXAMPLES   %%%
%%%%%%%%%%%%%%%%%%%%%%%%%%%%%%%%%%

\subsection{Numerical examples}
\label{subsec:1p1numerics}

In order to test the generalized harmonic shift condition we have
performed a series of numerical experiments.  We evolve Minkowski
initial data, but with a non-trivial initial slice given in Minkowski
coordinates $(t_M,x_M)$ as $t_M = p(x_M)$, with $p$ a profile function
that decays rapidly.  If we use $x=x_M$ as our spatial coordinate, the
spatial metric and extrinsic curvature turn out to be
\begin{equation}
g(t = 0) = 1 - {p^\prime}^2 \; , \qquad \tilde{K}(t=0) = - p'' / g \; .
\label{eq:1p1_initgKg}
\end{equation}
In all the simulations shown below we have taken for the function
$p(x)$ a Gaussian centered at the origin
\begin{equation}
p(x) = \kappa \; \exp \left[ - \left( \frac{x}{s} \right)^2 \right] \; .
\label{eq:1p1_inith}
\end{equation}
For our simulations we have chosen for $\kappa$ and $s$ the
same values used in~\cite{Alcubierre97a}, namely \mbox{$\kappa = 5$} 
and \mbox{$s = 10$}.  Furthermore, we start with unit lapse 
and vanishing shift.

All runs have been performed using a method of lines with fourth order
Runge-Kutta integration in time, and standard second order centered
differences in space.  Furthermore, we have used 64,000 grid points
and a grid spacing of $\Delta x = 0.0125$ (which places the boundaries
at $\pm 400$), together with a time step of $\Delta t = \Delta x/4$.
In the simulations shown below, we will concentrate on two different
aspects: First, we want to know how the generalized harmonic shift
condition works in practice, and what are its effects on the
evolution.  Also, we want to see if gauge shocks do form when they are
expected.

Furthermore, to study the overall growth in the evolution variables, we 
introduce the quantity $\delta$ defined through
\begin{equation}
\label{eq:definedelta}
\delta^2 := (\alpha - 1)^2 + \sigma^2 + (g - 1)^2 
+ \sum_{i=1}^4 v_i^2 \; ,
\end{equation}
as a measure of how non-trivial the data is.  For $\delta$ we then
also calculate the convergence factor $\eta$ which, using three runs
with high ($\delta^h$), medium ($\delta^m$) and low ($\delta^l$)
resolutions differing in each case by a factor of two, can be
calculated by
\begin{equation}
\eta = \frac{\frac{1}{N_i} \sum_{i = 1}^{N_i} | \delta^m_i - \delta^l_i |}
{\frac{1}{N_j} \sum_{j = 1}^{N_j} | \delta^h_j - \delta^m_j |} \; .
\end{equation}
In the plots we show three different convergence factors. In
particular, we denote with a triangle the convergence factor obtained
when comparing runs with $64,000$, $32,000$ and $16,000$ grid points
and a spatial resolution of $0.0125$, $0.25$ and $0.5$. We then use
boxes and diamonds when gradually lowering all three resolutions by a
factor of two.  For second order convergence we expect $\eta \simeq
4$.

\begin{figure}
\epsfxsize=85mm
\epsfbox{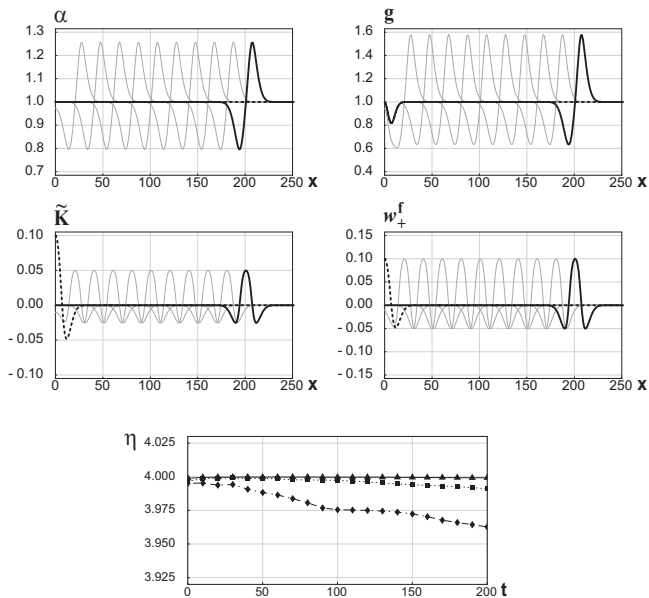}
\caption{For a simulation with harmonic slicing ($f=1$) and vanishing
shift, we show the evolution in time of the variables $\alpha$, $g$
and $\tilde{K}$, together with that of the eigenfield $w^f_+$.  The
values of the different quantities are shown every $\Delta t = 20$.
In the plot on the bottom we show three convergence factors when
increasing the resolution in the order ``diamond, box and triangle''.
In all three cases the convergence factor is close to the expected
value of 4..}
\label{fig:f1noshift}
\end{figure}

As a reference of what happens for the case of zero shift, in
Fig.~\ref{fig:f1noshift} we show a run that corresponds to harmonic
slicing and vanishing shift (these plots should be compared with
Fig. 2 of~\cite{Alcubierre97a}).  In the figures, the initial data is
shown as a dashed line, and the final values at $t=200$ as a solid
line (remember that the initial metric is non-trivial).  Intermediate
values at intervals of $\Delta t = 20$ are shown in light gray. As can
be seen from the plots, all variables behave in a wavelike fashion and
the convergence plot indicates that we have close to second order
convergence during the whole run for all resolutions considered.  Here
the pulses are moving out symmetrically in both directions away from
the origin (we only show the $x>0$ side).  One can see that the
initial non-trivial distortion in $g$ for small $x$ remains (so the
dashed and solid lines lie on top of each other there), indicating
that even though in the end we return to trivial Minkowski slices, we
are left with non-trivial spatial coordinates.

\begin{figure}
\epsfxsize=85mm
\epsfbox{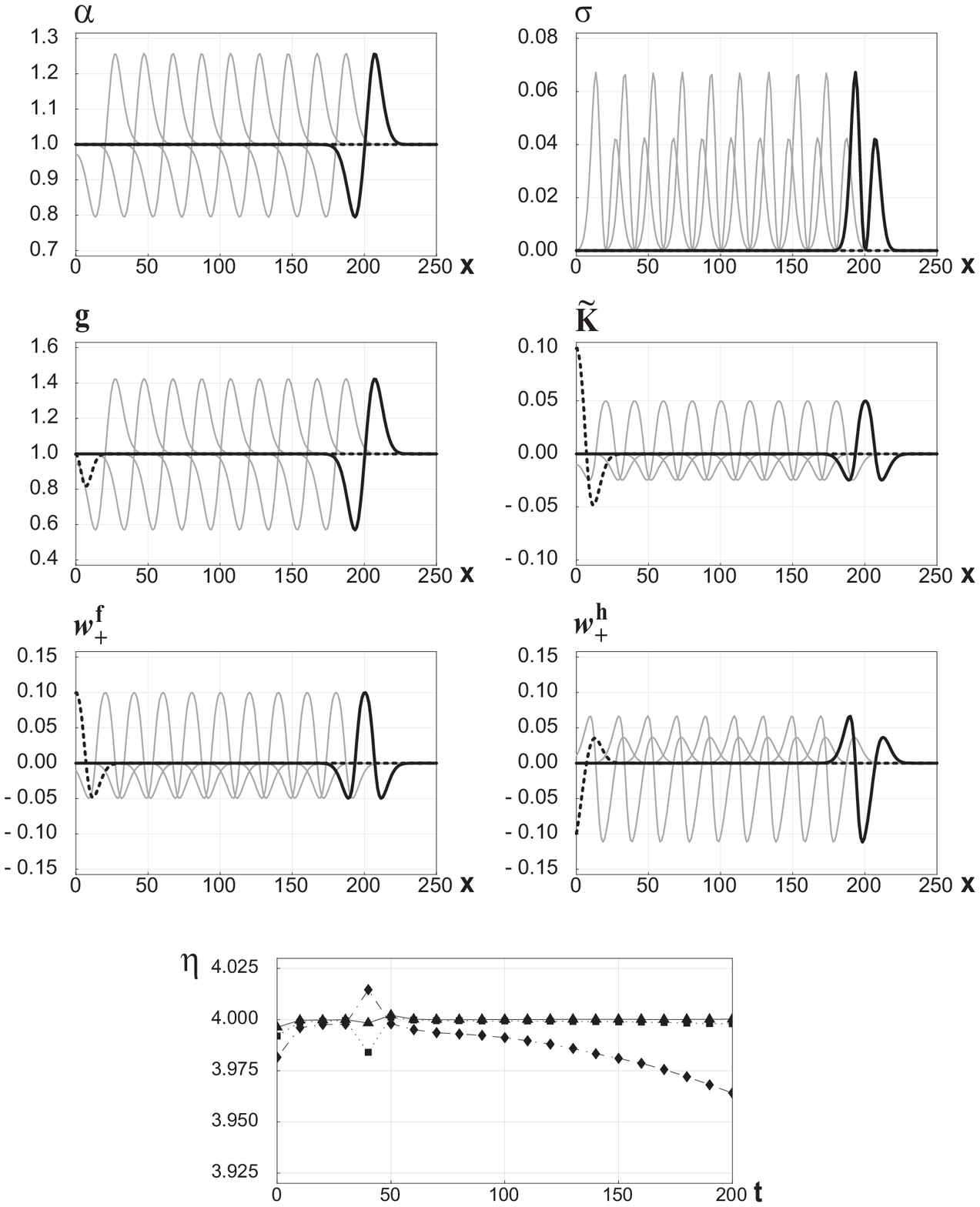}
\caption{For a simulation with harmonic slicing and harmonic shift
($f=h=1$), we show the evolution in time of $\alpha$, $\sigma$, $g$
and $\tilde{K}$, together with that of $w^f_+$ and $w^h_+$. As in the
previous figure, the bottom plot shows the convergence factors for
different resolutions.}
\label{fig:f1h1}
\end{figure}

Our second example is shown in Fig.~\ref{fig:f1h1} and corresponds to
$f=h=1$, that is, pure harmonic coordinates in both space and
time. The simulation is very similar to the previous one and
convergences again to second order.  The non-trivial shift, however,
behaves in such a way that at the end of the run no distortion remains
in the metric component $g$ at the origin. One should also note that
for \mbox{$f=h=1$} both eigenfields propagate with the same speed.
However, since quadratic and mixed source terms in the evolution
equations of both $w^f_\pm$ and $w^h_\pm$ are not present, simple
wavelike behavior for all variables is again observed.

This example allows us to understand the main effect that the
introduction of the generalized harmonic shift condition has on the
evolution: It drives the spatial coordinates to a situation where no
final distortion in the metric is present. In fact, it is not
difficult to understand why this is so.  From
equation~(\ref{eq:sigmadot_curve}) we can see that the sources for the
evolution of the rescaled shift are the derivatives of the lapse, the
trace of the extrinsic curvature and the $\Delta^l_{mn}$.  As the
shift condition does not feed back into the slicing condition (apart
from a trivial shifting of the time lines), the lapse and the trace of
the extrinsic curvature behave just as before, with pulses that
propagate away.  However, the $\Delta^l_{mn}$ will continue to drive
the evolution of the shift unless they become zero.  The behavior of
the shift condition is then to drive the system to a situation where
the $\Delta^l_{mn}$ vanish.  In the simple 1+1 case this is equivalent
to reaching a state where the spatial metric itself becomes trivial.

\begin{figure}
\epsfxsize=85mm
\epsfbox{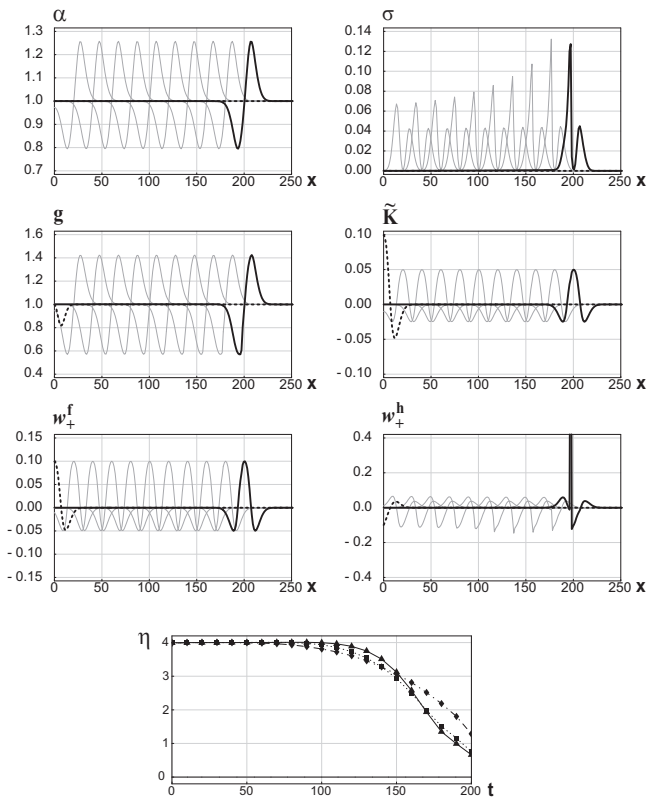}
\caption{For $f=1$ and $h = 1 + 3\sigma^2$, the simulation fails
shortly before the time $t=200$ due to a sharp gradient developing in
the rescaled shift $\sigma$, and a corresponding large spike appearing
in the eigenfield $w^h_\pm$.  The bottom plot shows that convergence
starts to be lost at $t \simeq 150$ (notice the change of scale as 
compared to previous plots), indicating that a blow-up has
happened at around this time. This type of behavior is expected in
this case since the source criteria is not satisfied.}
\label{fig:f1hsigma}
\end{figure}

A third example is presented in Fig.~\ref{fig:f1hsigma}, which uses
again $f = 1$, but now we take $h = 1 + 3\sigma^2$.  Initially, the
evolution behaves in a very similar way to the previous case.  At
later times, however, we observe that a sharp gradient develops in the
rescaled shift $\sigma$, with a corresponding large spike in the shift
eigenfield $w^h_\pm$.  Moreover, from the convergence plot we see that
there is a clear loss of convergence, and as the resolution is
increased, this loss of convergence becomes more sharply centered
around a specific time $t \simeq 150$, indicating that a blow-up
happens at this time.  Since in this case $h$ is a function of
$\sigma$, the source criteria is clearly not satisfied.  The fact that
a large spike has developed in $w^h_\pm$ therefore strengthens the
case for the source criteria being a good indicator of when blow-ups
can be expected.

\begin{figure}
\epsfxsize=85mm
\epsfbox{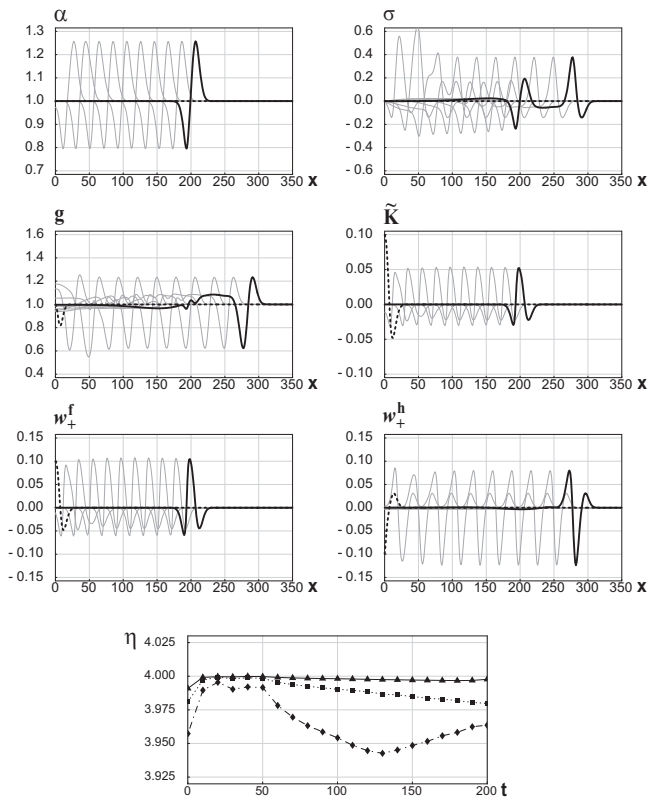}
\caption{For $f=1$ and $h=2$, the lapse and shift eigenfields travel
at different speeds. The lapse, extrinsic curvature and eigenfield
$w^f_+$ show a pulse traveling with roughly unit speed, while the
eigenfield $w^h_+$ shows a pulse moving with speed $\sim \sqrt{2}$.
The metric $g$ and rescaled shift $\sigma$, on the other hand,
separate into two pulses traveling at the two different eigenspeeds.}
\label{fig:f1h2}
\end{figure}

Our final example uses $f=1$ and $h=2$, and is shown in
Fig.~\ref{fig:f1h2}.  This example is interesting as the speeds for
the lapse and shift eigenfields are different.  Concentrate first on
the evolution of the lapse, the extrinsic curvature and $w^f_+$.  Here
the evolution is essentially identical to that of the previous
examples, converging again to second order: a pulse travels with
roughly unit speed and behind it everything rapidly relaxes back to
trivial values.  The eigenfield $w^h_+$, on the other hand, shows a
pulse traveling faster, with a speed $\sim\sqrt{2}$. It also takes
considerably longer for the region behind this pulse to relax to
trivial values.  Finally, the metric $g$ and rescaled shift $\sigma$
separate into two pulses traveling at the two different eigenspeeds.
This is to be expected, as from~(\ref{eq:Dsigma_eigen})
and~(\ref{eq:Dg_eigen}) we see that metric and shift have
contributions from both types of eigenfields.

\begin{figure}[ht!]
\epsfxsize=85mm
\epsfbox{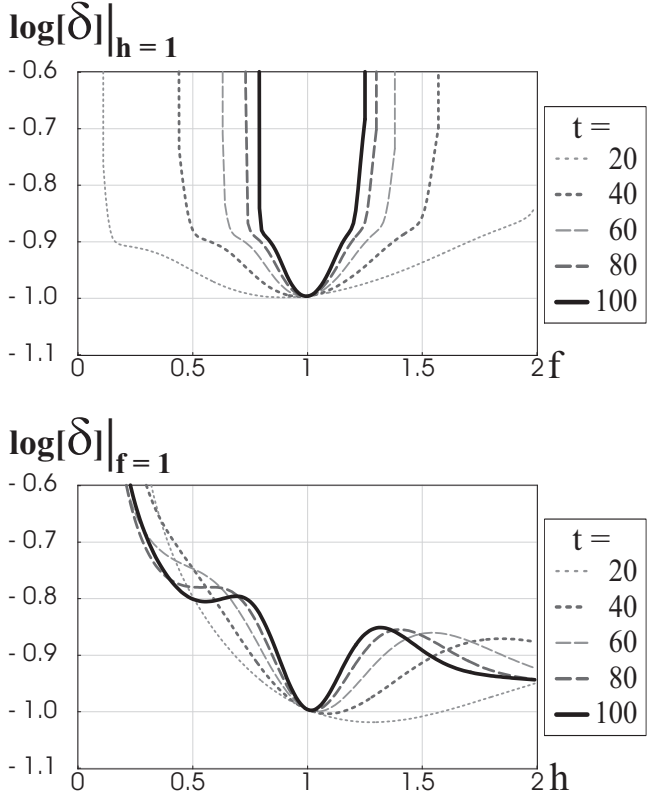}
\caption{\textit{Top.} For evolutions with $h=1$, the rms of $\delta$
is shown on a logarithmic scale as a function of $f$ every $\Delta t =
20$.  The value $f =1$ is obviously preferred. \textit{Bottom.} For
runs with harmonic slicing ($f=1$), the same quantity is plotted as a
function of $h$.  Here $h=1$ is the optimal choice, but $h \sim 0.5$ or 
$h \gg 1$ is also preferred.}
\label{fig:deltaoffandh}
\end{figure}

For the different runs we have also studied the behavior of the
logarithm of the root mean square (rms) of $\delta$ over time.  Since
the behavior of the evolution turns out to depend to some extend on
the initial data, and in particular on the sign of the Gaussian in
(\ref{eq:1p1_inith}), we perform runs for both $\kappa = 5$ and
$\kappa = -5$, and then take the average of both runs when calculating
$\delta$.  For the initial data we are using, at time $t=0$ this
yields a value ${\rm log} (\delta) \approx -1.583$ for both signs of
$\kappa$. In Fig.~\ref{fig:deltaoffandh} we plot the rms of the
quantity $\delta$ for the times $t = \{ 20, 40, 60, 80,1 00 \}$, when
using either $h = 1$ and varying the (constant) value of $f$ (top
panel), or using $f = 1$ together with different (again constant)
values of $h$ (bottom panel).  From the top panel we see that $f=1$ is
clearly preferred.  In addition we want to point out that runs with $f
< 0.79$ and $f > 1.25$ crashed before reaching the time $t =
100$. This behavior is expected as we know that constant values of $f$
different from one produce blow-ups.  In the lower panel we observe
that for $f=1$ corresponding to harmonic slicing, $h=1$ performs best.
In addition, values $h \sim0.5$ and $h \gg 1$ also seem to be
preferred.  One should note that mixed terms $w^f_\pm w^h_\pm$ in the
evolution equations of both $w^f_\pm$ and $w^h_\pm$ for these choices
of $h$ play a minor role since localized perturbations in these
eigenfields separate quickly when traveling with different speeds.  We
also want to mention that for $h < 0.19$ the simulations again crashed
before reaching the time $t=100$.  The observation that $\delta$ grows
rapidly and runs crash early if $f$ and/or $h$ are very close to zero
can be understood by the fact that the system is not strongly
hyperbolic if $f=0$ and/or $h=0$.

\begin{figure}[hb!]
\epsfxsize=85mm
\epsfbox{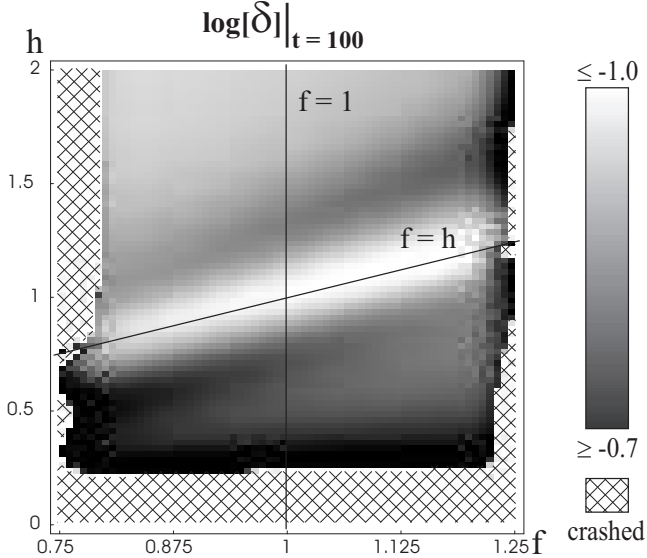}
\caption{Contour plot of the rms of $\delta$ at time $t = 100$.  Small
values for this quantity are found for $f = 1$, when $h = 1$ or $h \gg
1$, and also for $f = h$.}
\label{fig:hfcontour_1p1}
\end{figure}

In the contour plot of Fig.~\ref{fig:hfcontour_1p1} we show the rms of
$\delta$ at time $t = 100$ as a function of the gauge parameters $f$
and $h$, using $64 \times 80$ equidistant parameter choices.  Cases
that have already crashed by that time correspond to the hashed
regions.  Note that the darker regions in this plot denote parameter
choices where a significant growth in the evolution variables is
present, while brighter regions correspond to runs with very little
growth.  We find small values for the rms of $\delta$ for $f$ being
close to its shock avoiding value $f=1$, and either $h=1$ or $h \gg
1$.  In addition, we can also observe that $f = h$ corresponds to a
preferred choice.  This can be explained by the fact that for this
gauge choice the mixed terms $w^f_\pm w^h_\pm$ are missing in the
evolution equations of both $w^f_\pm$ and $w^h_\pm$.

%%%%%%%%%%%%%%%%%%%%%%%%%%%%%%%%%%%%%%%%
%%%   SPHERICALLY SYMMETRIC SYSTEM   %%%
%%%%%%%%%%%%%%%%%%%%%%%%%%%%%%%%%%%%%%%%

\section{Einstein equations in spherical symmetry}
\label{sec:sphsymm}

As a second application of the generalized harmonic shift condition we
will consider vacuum general relativity in spherical symmetry.  This
situation is considerably richer than the 1+1 dimensional case, but it
also presents some special problems because of the singular nature of
spherical coordinates at the origin.

%%%%%%%%%%%%%%%%%%%%%%%%%%%%%%%%%%%%%%%%%
%%%   SPHERICAL SYMMETRY: EQUATIONS   %%%
%%%%%%%%%%%%%%%%%%%%%%%%%%%%%%%%%%%%%%%%%

\subsection{ADM evolution equations}
\label{subsec:sphsymmequations}

We will consider the spherically symmetric line element written in the
form
\begin{eqnarray} 
ds^2 &=& - \alpha^2 \left( 1 - A \sigma^2 \right) dt^2 
+ 2 \alpha A \sigma dr dt \nonumber \\
&& + A dr^2 + B r^2 d\Omega^2 \; ,
\label{eq:sphericalmetric} 
\end{eqnarray}
where all the metric coefficients are functions of both $t$ and $r$.
We now introduce the following auxiliary variables
\begin{eqnarray}
D_{\alpha} := \partial_r \ln \alpha \; , & \qquad &
d_{\sigma} := \partial_r \sigma \; , 
\label{eq:defDalphadsigma}\\
D_A := \partial_r \ln A \; , & \qquad &
D_B := \partial_r \ln B \; .
\label{eq:defDADB}
\end{eqnarray}
Notice again that we use logarithmic derivatives for the lapse and the
spatial metric, but only an ordinary derivative for the shift.  For
the extrinsic curvature, we will use the mixed components
\begin{equation}
K_A := K_r^r \; , \qquad
K_B := K^\theta_\theta = K^\phi_\phi \; .
\label{eq:defKAKB}
\end{equation}
Following~\cite{Alcubierre04a}, we will change our main evolution
variables and make use of the ``anti-trace'' of the metric spatial
derivatives \mbox{$D = D_A - 2D_B$}, and the trace of the extrinsic
curvature \mbox{$K = K_A + 2K_B$}, instead of $D_A$ and $K_A$.

For the regularization of the evolution equations at the origin we
will follow the procedure described in~\cite{Alcubierre04a}, which
requires the introduction of an auxiliary variable
\begin{equation}
\lambda = (1 - A/B)/r \; .
\label{eq:lambda}
\end{equation}
Local flatness guarantees that $\lambda$ is regular and of order $r$
near the origin.  By taking $\{ \alpha, A, B, d_\sigma, K, K_B \}$ as
even functions at $r=0$, and $\{ \sigma, D_\alpha, D, D_B, \lambda \}$
as odd, one obtains regular evolution equations at $r=0$.

In terms of the variables introduced above, the hamiltonian and
momentum constraints become (in vacuum)
\begin{eqnarray}
0 &=& {\cal C}_h =  - \partial_r D_B + \frac{D_B}{2} \:
\left( D + \frac{D_B}{2} \right) \\
&+& A K_B ( 2 K - 3 K_B) 
+ \frac{1}{r} \left( D - D_B - \lambda \right) \; , 
\label{eq:ham} \nonumber \\
0 &=& {\cal C}_m = - \partial_r K_B + (K - 3 K_B) \left[ \frac{D_B}{2}
+ \frac{1}{r} \right] \; . \hspace{5mm} 
\label{eq:mom}
\end{eqnarray}
Notice that the hamiltonian constraint is regular, while the momentum
constraint still has the term \mbox{$(K - 3 K_B)/r \equiv (K_A -
K_B)/r$} which has to be handled with care numerically.  This is not a
problem as the momentum constraint does not feed back into the ADM
evolution equations. On the other hand, when one adds multiples of the
momentum constraint to the evolution equations in order to obtain
strongly hyperbolic re-formulations (as in the following section), the
regularization procedure requires some of the dynamical variables to
be redefined by adding to them a term proportional to $\lambda$
(see~\cite{Alcubierre04a} for details).  This redefinition, however,
does not affect the characteristic structure of the system.  Because
of this, in the following analysis we will simply ignore this issue.

For the evolution of the lapse we will again take the Bona-Masso
slicing condition, which in spherical symmetry takes the form
\begin{equation}
\partial_t \alpha = 
\alpha^2 \left( \sigma D_\alpha - f K \right) \: .
\label{eq:alphadot}
\end{equation}
For the shift we will use the generalized harmonic shift condition in
the form~(\ref{eq:sigmadot_curve}).  In this case one finds
\begin{eqnarray}
\gamma^{mn} \: {}^{(3)}\Gamma^r_{mn} &=& \frac{D}{2 A} - \frac{2}{r A} \; , \\
\left. \gamma^{mn} \: {}^{(3)}\Gamma^r_{mn} \right|_{\rm flat}
&=& - \frac{2}{r B} \; .
\end{eqnarray}
Notice that in the first of these expressions we have used the
Christoffel symbols for the full spatial metric \mbox{$dl^2 = A dr^2 +
B r^2 d \Omega^2$}, while in the second we used those of the flat
metric \mbox{$dl^2 = dr^2 + r^2 d \Omega^2$}.  However, as is clear
from~(\ref{eq:sigmadot_curve}), in both cases we have to contract indices 
using the full inverse metric which explains why there is a factor $B$ in
the denominator of the second expression.  Using these expressions we
then find
\begin{equation}
\Delta^r = \frac{D}{2 A} - \frac{2 \lambda}{A} \; ,
\end{equation}
with $\lambda$ defined in~(\ref{eq:lambda}) above.  Our final shift
condition is then regular at the origin and has the form
\begin{equation}
\partial_t \sigma = \alpha \left[ \sigma d_\sigma
- \frac{D_\alpha}{A} + h \left( \frac{D}{2 A} + \sigma K
- \frac{2 \lambda}{A} \right) \right] \: .
\label{eq:sigmadot}
\end{equation}
It is important to mention that if we had used the original
condition~(\ref{eq:sigmadotGH}) instead of~(\ref{eq:sigmadot_curve}),
we would have found that the shift evolution equation was singular.
Moreover, one also finds that taking $\sigma^\theta$ and $\sigma^\phi$
equal to zero is consistent when using~(\ref{eq:sigmadot_curve}) in
the sense that their respective evolution equations guarantee that
they remain zero, which would not have been the case
with~(\ref{eq:sigmadotGH}).

Going back to the metric components $A$ and $B$, we find for their
evolution equations
\begin{eqnarray}
\partial_t A &=& 2 \alpha A \left[ 
\sigma \left( D_\alpha + \frac{D}{2} + D_B \right) 
\right. \nonumber \\
&+& \left. d_\sigma - K + 2K_B \rule{0mm}{5mm} \right] \; , 
\label{eq:Adot} \\ 
\partial_t B &=& 2 \alpha B \left[ 
\sigma \left( \frac{D_B}{2} + \frac{1}{r} \right) - K_B \right] \; . 
\label{eq:Bdot}
\end{eqnarray}

The evolution equations for $D_\alpha$, $d_\sigma$, $D$ and $D_B$
again follow trivially from the above equations.  Finally, the ADM
evolution equations for the extrinsic curvature components turn out to
be
\begin{eqnarray}
\partial_t K &=& \frac{\alpha}{A} \left\{ \rule{0mm}{5mm} - \partial_r D_\alpha
- 2 \: \partial_r D_B + \sigma A \: \partial_r K \right. \nonumber \\
&+& D_\alpha \left( \frac{D}{2} - D_\alpha \right) 
+ D_B \left( D + \frac{D_B}{2} \right) \nonumber \\ 
&+& \left. A K^2 - \frac{2}{r} \left( D_\alpha - D + D_B 
+ \lambda \right) \right\} \: ,
\label{eq:trKdot} \\
\partial_t K_B &=&  \frac{\alpha}{A} \left\{ - \frac{\partial_r D_B}{2} 
+ \sigma A \: \partial_r K_B - \frac{D_{\alpha} D_B}{2} 
+ \frac{D D_B}{4}  \right. \nonumber \\
&+& \left. A K K_B -\frac{1}{r} \left( D_\alpha - \frac{D}{2} + D_B 
+ \lambda \right) \right\} \: .
\label{eq:KBdot}
\end{eqnarray}
Notice that these are directly the standard ADM evolution equations
written in terms of $\{K,K_B\}$, with no multiples of the constraints
added to them.  In the next section we will consider how such
adjustments affect the hyperbolicity of the full system.

%%%%%%%%%%%%%%%%%%%%%%%%%%%%%%%%%%%%%%%%%%%%%
%%%   SPHERICAL SYMMETRY: HYPERBOLICITY   %%%
%%%%%%%%%%%%%%%%%%%%%%%%%%%%%%%%%%%%%%%%%%%%%

\subsection{Adjustments and hyperbolicity}
\label{subsec:sphsymhyp}

In order to analyze the characteristic structure of the full system of
evolution equations including the gauge conditions, we start by
defining
\begin{eqnarray}
u &:=& \left( \alpha, \sigma, A, B, \lambda \right) \; , \\
v &:=& \left( D_\alpha, d_\sigma, D, D_B, K, K_B \right) \; .
\end{eqnarray}
The system of equations can then be written in the form
(\ref{eq:uPDE})-(\ref{eq:vPDEmatrix}).  It turns out that by doing
this, one finds that the ADM evolution system introduced above is not
strongly hyperbolic when $f=1$ and/or $h=1$.  This is undesirable, as
these cases correspond precisely to purely harmonic coordinates.

Following~\cite{Reimann:2004wp}, in order to obtain strongly
hyperbolic systems we will consider adjustments to the evolution
equations of the extrinsic curvature components $K$ and $K_B$ of the
form
\begin{equation}
\partial_t v_i + \sum_{j=1}^m A_{ij} \partial_r v_j 
+ h_i \frac{\alpha}{A} \: C_h = q_i \; .
\end{equation}

Note that we are considering only very restricted adjustments here.
In particular, we do not modify the evolution equations for the $D$'s
and for $d_\sigma$.  As explained in Ref.~\cite{Reimann:2004wp}, this
is important for the blow-up analysis in the next section, as
otherwise the constraints that link the $D$'s to derivatives of the
$u$'s will fail to hold and the analysis breaks down~\footnote{It is
in fact possible to lift this restriction, but the analysis would
become more involved.  Since we are mostly interested in studying the 
effects of the generalized harmonic shift condition here we prefer not to 
complicate the analysis any further.}.
Furthermore, for simplicity we will not consider adjustments that use
the momentum constraint.  

For the coefficients $h_K$ and $h_{K_B}$ we make the following ansatz
\begin{eqnarray}
 h_K &=& - 2 + b(\alpha,\sigma,A,B) \; ,
\label{eq:hKansatz} \\
 h_{K_B} &=& \left[ c(\alpha,\sigma,A,B) - 1 \right] / 2 \; .  
\label{eq:hKBansatz}
\end{eqnarray}
With these adjustments we find that the characteristic matrix for our
system of evolution equations becomes
\begin{equation}
{\bf A} = \alpha \left( \begin{array}{cccccc}
- \sigma  & 0        & 0        & 0        & f          & 0 \\
1/A       & - \sigma & - h / 2A & 0        & - h \sigma & 0 \\ 
-2 \sigma & - 2      & -\sigma  & 0        & 2          & -8 \\
0         & 0        & 0        & - \sigma & 0          & 2 \\
1/A       & 0        & 0        & b / A    & - \sigma   & 0 \\
0         & 0        & 0        & c / 2A   & 0          & - \sigma
\end{array} \right) .
\label{eq:A_ss}
\end{equation}
One may now readily verify that this matrix has the following
eigenvalues
\begin{eqnarray}
\lambda^f_\pm & = & \alpha \left( - \sigma \pm \sqrt{f/A} \right) \; , \\
\lambda^h_\pm & = & \alpha \left( - \sigma \pm \sqrt{h/A} \right) \; , \\
\lambda^c_\pm & = & \alpha \left( - \sigma \pm \sqrt{c/A} \right) \; .
\end{eqnarray}
The system is therefore hyperbolic for $\{ f,h,c \} > 0$.
Furthermore, there exists a complete set of eigenvectors as long as $c
\neq f$ and $c \neq h$, so the system is strongly hyperbolic except in
those two cases. The eigenfields turn out to be:
\begin{eqnarray}
w^f_\pm & = & \left( c-f \right) D_\alpha - b f D_B \nonumber \\
&\pm& \sqrt{f A} \left[ \left( c-f \right) K - 2b K_B \right] \; , \\
w^h_\pm & = & \left( c - h \right) \left\{
A^{1/2} \left[ d_\sigma - \left( 1 \pm \sigma \sqrt{h A} \right) K \right]
\mp \sqrt{h} \: \frac{D}{2} \right\} \nonumber \\
&\pm& \sqrt{h} \left[ b \left( 1 \pm \sigma \sqrt{h A} \right) 
- 2c \right] D_B \nonumber \\
&+& 2 \sqrt{A} \left[ b \left( 1 \pm \sigma \sqrt{h A} \right) 
- 2h \right] K_B \; , \\
w^c_\pm & = & \sqrt{c} \: D_B \pm 2 \sqrt{A} \: K_B \; .
\end{eqnarray}
It is clear from these expressions that when $c=f$ the first and third
pairs of eigenfields become proportional to each other and are
hence no longer independent, while for $c=h$ it is the second and
third pairs that become proportional.

%%%%%%%%%%%%%%%%%%%%%%%%%%%%%%%%%%%%%%%%%%%%%%
%%%   SPHERICAL SYMMETRY: SHOCK ANALYSIS   %%%
%%%%%%%%%%%%%%%%%%%%%%%%%%%%%%%%%%%%%%%%%%%%%%

\subsection{Gauge and constraint shocks}
\label{subsec:sphsymshocks}

As we did for the 1+1 dimensional system, we will now study the 
possible formation of blow-ups for the evolution equations in spherical
symmetry. In order to apply the source criteria for avoiding blow-ups
we need to calculate the quadratic source terms in the evolution
equations for the eigenfields.  We first look for ``gauge shocks'',
for which we concentrate on the gauge eigenfields $w^f_\pm$ and
$w^h_\pm$.  For the quadratic source terms we find
\begin{eqnarray}
c^{fff}_{\pm\pm\pm} &\propto& \frac{1}{(c-f)} \; 
\left( 1 - f - \frac{\alpha f'}{2} \right) \; , \\
c^{hhh}_{\pm\pm\pm} &\propto& \frac{1}{(c-h)} \;
\frac{\partial h}{\partial \sigma} \; .
\end{eqnarray}
Demanding now that these terms vanish we obtain precisely the same
conditions on $f$ and $h$ as in the 1+1 dimensional case.  So again 
$f=1 + {\rm const}/\alpha^2$ and $h=h(\alpha)$ are shock avoiding
solutions.  Furthermore, if one chooses $f = h = 1 + {\rm
const}/\alpha^2$, mixed terms of the form $w^f_\pm w^h_\pm$ do not
appear in the evolution equations of the gauge eigenfields $w^f_\pm$
and $w^h_\pm$, so this is a preferred choice.

In contrast to the 1+1 dimensional case, now also blow-ups associated 
with the constraint eigenpair $w^c_\pm$ can arise.  The quadratic 
coefficient in this case takes the form
\begin{equation}
c^{ccc}_{\pm\pm\pm} \propto \left( 1 - 4b + 3c \right) \; ,
\end{equation}
and by asking for this coefficient to vanish we find
\begin{equation}
b = \left( 1 + 3c \right)/ 4 \; .
\label{eq:bofc}
\end{equation}
From~(\ref{eq:hKansatz}) and~(\ref{eq:hKBansatz}) we then infer that
$h_K$ and $h_{K_B}$ are related by
\begin{equation}
h_K = -1 + \frac{3 \, h_{K_B}}{2} \; , \qquad h_{K_B} > - \frac{1}{2} \; ,
\label{eq:hKhKBrelation}
\end{equation}
which is the precisely the same constraint shock avoiding half-line in
the $\{ h_K, h_{K_B} \}$ parameter space that was found in
Ref.~\cite{Reimann:2004wp}.

%%%%%%%%%%%%%%%%%%%%%%%%%%%%%%%%%%%%%%%%%%%%%%%%%%
%%%   SPHERICAL SYMMETRY: NUMERICAL EXAMPLES   %%%
%%%%%%%%%%%%%%%%%%%%%%%%%%%%%%%%%%%%%%%%%%%%%%%%%%

\subsection{Numerical examples}
\label{sec:sphsymmnumerics}

We will now test the effects of the generalized harmonic shift in
spherical symmetry by performing a series of numerical simulations.
As in the 1+1 dimensional case, we will concentrate on two aspects,
namely the effect of the shift condition on the evolution of the
metric, and the possible formation of blow-ups. Furthermore, in order
to decouple geometric effects associated with the center of symmetry,
we will consider two distinct regimes, one far from the origin and one
close to it.

%%%%%%%%%%%%%%%%%%%%%%%%%%
%%%   MINKOWSKI: FAR   %%%
%%%%%%%%%%%%%%%%%%%%%%%%%%

\subsubsection{Pulses far from the origin}
\label{subsubsec:Minkowskifar}

We will first consider simulations that are far from the origin, using
initial data that is similar to the one used in
Sec.~\ref{subsec:1p1numerics}.  We start with the Minkowski spacetime, 
but use a non-trivial initial slice with a profile $t_M=p(r_M)$.  The
initial metric and extrinsic curvature then become
\begin{eqnarray}
A(t=0) &=& 1 - p'^2 \; , \label{eq:sphsymm_initA} \\
B(t=0) &=& 1 \; , \label{eq:sphsymm_initB} \\
K(t=0) &=& - \frac{1}{\sqrt{A}} 
\left( \frac{p''}{A} + \frac{2 p'}{r} \right) \; ,
\label{eq:sphsymm_initK} \\
K_B(t=0) &=& - \frac{p'}{r \sqrt{A}} \; .
\label{eq:sphsymm_initKB}
\end{eqnarray}
The profile function $p(r)$ is again chosen to be a Gaussian,
\begin{equation}
p(r) = \kappa \; \exp 
\left[ - \left( \frac{r - r_c}{s} \right)^2 \right] \; ,
\label{eq:sphsymm_inith}
\end{equation}
using for its amplitude and width the values \mbox{$\kappa = \pm 5$}
and \mbox{$s = 10$}.  The center of the Gaussian is taken at
$r_c=250$, such that for evolution times of $t \sim 100$ the
perturbation will remain away from the origin.

We have performed runs with the code described
in~\cite{Alcubierre04a}, which uses a method of lines with fourth
order Runge-Kutta integration in time, and standard second order
centered differences in space.  We used 5,000 grid points and a grid
spacing of $\Delta r = 0.1$ (which places the outer boundary at $500$)
together with a time step of $\Delta t = \Delta r/4$.

\begin{figure}
\epsfxsize=85mm
\epsfbox{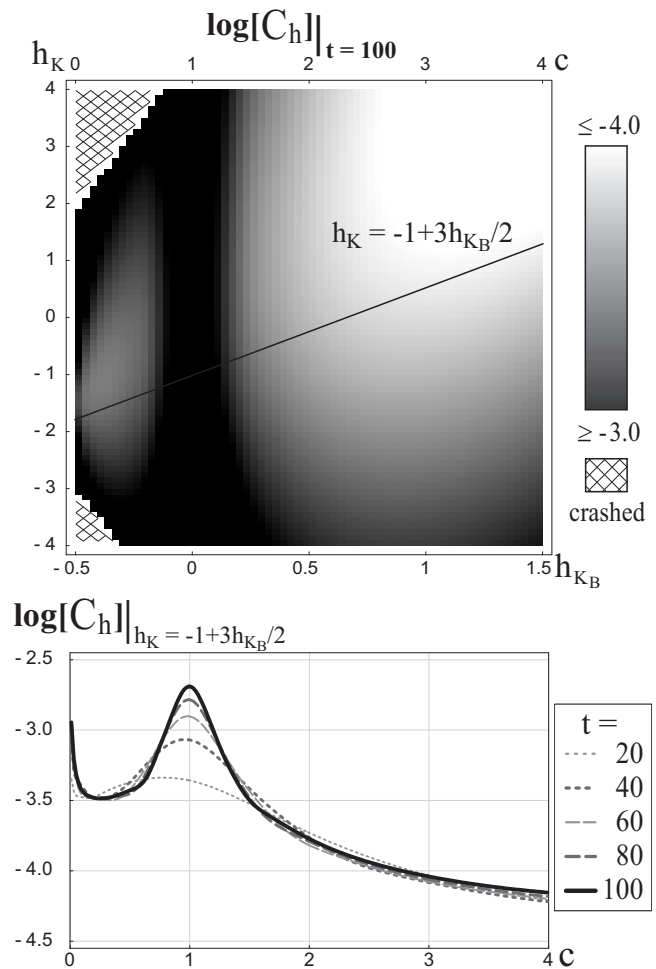}
\caption{\textit{Top.} Contour plot of the rms of the hamiltonian
constraint at time $t = 100$ as a function of the adjustment
parameters $h_K$ and $h_{K_B}$. The parameter line suggested by the
source criteria, $h_K = -1 + 3 h_{K_B}/2$ with $h_{K_B} > -1/2$, is
shown as a solid line.  \textit{Bottom.} For reasons described in the
text, along this line $c \sim 1/4$ and $c \gg 1$ are preferred values
for $c$, the latter determining the eigenspeeds $\lambda^c_\pm$ of the
constraint modes.}
\label{fig:constraints}
\end{figure}

In a preparatory experiment, we studied which evolution systems
perform best for our optimal gauge choice $f=h=1$.  The upper panel of
Fig.~\ref{fig:constraints} (which should be compared with Fig.~7
of~\cite{Reimann:2004wp}) shows the rms of the hamiltonian constraint
at time $t=100$, as a function of the adjustment parameters $h_K$ and
$h_{K_B}$.  We can see that the line~(\ref{eq:hKhKBrelation}) obtained
by the source criteria is numerically preferred, although there does
seem to be a discrepancy for large values of $h_{K_B}$ for which the
numerical results suggest a somewhat steeper line.  This discrepancy
is due to the effect of $1/r$ terms which are not taken into account
by the source criteria and can be eliminated by removing these terms
by hand.  It is also important to point out that, in contrast to
Ref.~\cite{Reimann:2004wp}, the initial data used here satisfies the
constraints and all subsequent constraint violations are caused by
truncation error.

In order to determine which points on this line perform best, {\em
i.e.} to fix the eigenspeeds $\lambda^c_\pm$ of the constraint mode,
we tested different (constant) values of $c$.  From the lower panel of
Fig.~\ref{fig:constraints} (to be compared with Fig.~5 of
\cite{Reimann:2004wp}) we find that values $c \sim 1/4$ and $c \gg 1$
are preferred.  This observation can be readily understood by the fact
that the system is not strongly hyperbolic for $c = 0$ and $c = 1$,
and by the fact that for $c \sim 1$ we expect contributions from mixed
source terms, since then $w^f_\pm$, $w^h_\pm$ and $w^c_\pm$ propagate
with similar or even identical eigenspeeds.

\begin{figure*}
\epsfxsize=180mm
\epsfbox{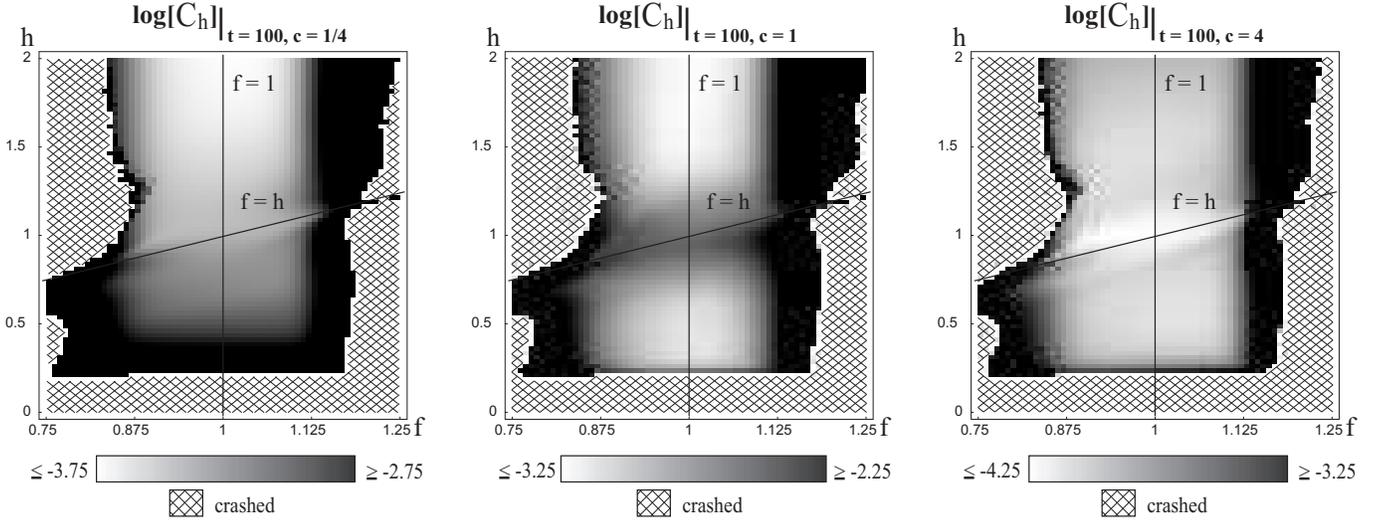}
\caption{Contour plot of the rms of the hamiltonian constraint at time
$t = 100$ for $c=\{ 1/4,1,4 \}$.  As in the 1+1 dimensional case,
$f=1$ together with $h=1$ or $h \gg 1$, and $f=h$ are preferred
parameter choices.  Furthermore, by far the lowest constraint
violations are found when the constraint eigenspeed is different from
the gauge eigenspeeds, {\em i.e.} for $c \ne 1$.}
\label{fig:hfcontour_sph}
\end{figure*}

For our main experiment regarding gauge effects, we concentrated on
evolution systems which belong to the shock avoiding family \mbox{$h_K
= -1 + 3 h_{K_B}/2$}, where for $c$ we considered three different
values: $c = \{ 1/4,1,4 \}$.  As long as the pulses remain far from
the origin, we have found that the evolutions behave in a very similar
way to those of the 1+1 dimensional case described in
Sec.~\ref{subsec:1p1numerics}.  We summarize these results in
Fig.~\ref{fig:hfcontour_sph}, showing for these three choices of $c$
the rms of the hamiltonian constraint at time $t = 100$ as a function
of $f$ and $h$.  These graphs are very similar to
Fig.~\ref{fig:hfcontour_1p1} and show that $f=1$ together with $h=1$
or $h \gg 1$, and $f=h$ are again preferred parameter choices,
indicating that the same mechanisms as in the 1+1 dimensional case are
at work.  One should observe the different scales when comparing the
three plots corresponding to different values of $c$, which indicate
that by far the lowest constraint violations are found when the
constraint eigenspeed is different from the gauge eigenspeeds.  Notice
also in the middle plot corresponding to $c=1$ that the region around 
$f=h=1$ is in fact dark.  This can be explained by the fact that for 
$f=h=c=1$ the evolution system is not strongly hyperbolic.

When the pulses come close to the origin, however, additional effects
arise due to $1/r$ terms.  In the next section we will consider this
situation.

%%%%%%%%%%%%%%%%%%%%%%%%%%%%
%%%   MINKOWSKI: CLOSE   %%%
%%%%%%%%%%%%%%%%%%%%%%%%%%%%

\subsubsection{Pulses close to the origin}
\label{subsubsec:Minkowskiclose}

In order to see directly the effect of the generalized harmonic shift
condition on the evolution of the geometric variables, we will
consider again a series of simulations of Minkowski spacetime, but
this time close to the origin $r=0$.  The initial data for these runs
will be simpler than the one used in the previous section: We start
with a flat Minkowski slice with $A=B=1$ and $K_A = K_B = 0$, and take
a non-trivial initial lapse of the form
\begin{equation}
\alpha = 1 + \kappa r^2  \left( e^{- (r - r_c)^2/s^2}
+ e^{- (r + r_c)^2/s^2} \right) \; ,
\end{equation}
with $\kappa = 10^{-5}$, $r_c = 10$ and $s = 1$ (the reason for the
two gaussians is to make sure the initial lapse is an even function of
$r$). All simulations shown here use 4,000 grid points, with a grid
spacing of $\Delta r = 0.01$ (which places the outer boundary at
$r=40$), together with a time step of $\Delta r/4$. In the plots, the
initial data is shown as a dashed line and the final values at $t=20$
as a solid line.  Intermediate values are plotted every $\Delta t = 2$
in light gray.

\begin{figure}
\epsfxsize=70mm
\epsfbox{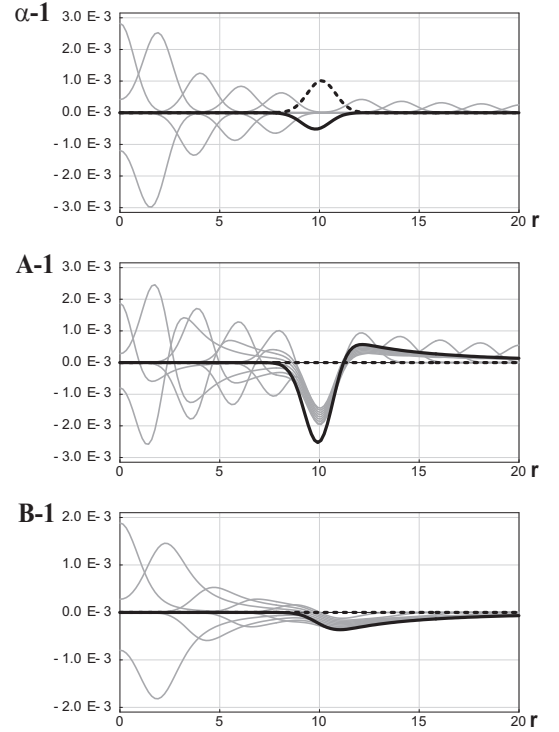}
\caption{Evolution of pulses close to the origin for harmonic slicing
($f=1$) and zero shift. In order to make the details more visible,
here we plot $\alpha-1$, $A-1$ and $B-1$.}
\label{fig:sphsymmnoshift}
\end{figure}

As reference, we first show in Fig.~\ref{fig:sphsymmnoshift} a run for
the case of harmonic slicing ($f=1$) with no shift. In order to look
at the details in a clearer way, in the figure we plot $\alpha-1$,
$A-1$ and $B-1$. As expected, the perturbation pulse in the lapse
separates into two pulses, one moving outward and one inward. The
inward moving pulse goes through the origin and starts moving out much
in the way a simple scalar wave would.  The pulses in the lapse are
accompanied by similar pulses in the metric variables $A$ and $B$.
However, one can clearly see that the metric variables are not
evolving toward trivial values, so in the end we are left with
Minkowski slices with non-trivial spatial coordinates.

\begin{figure}
\epsfxsize=70mm
\epsfbox{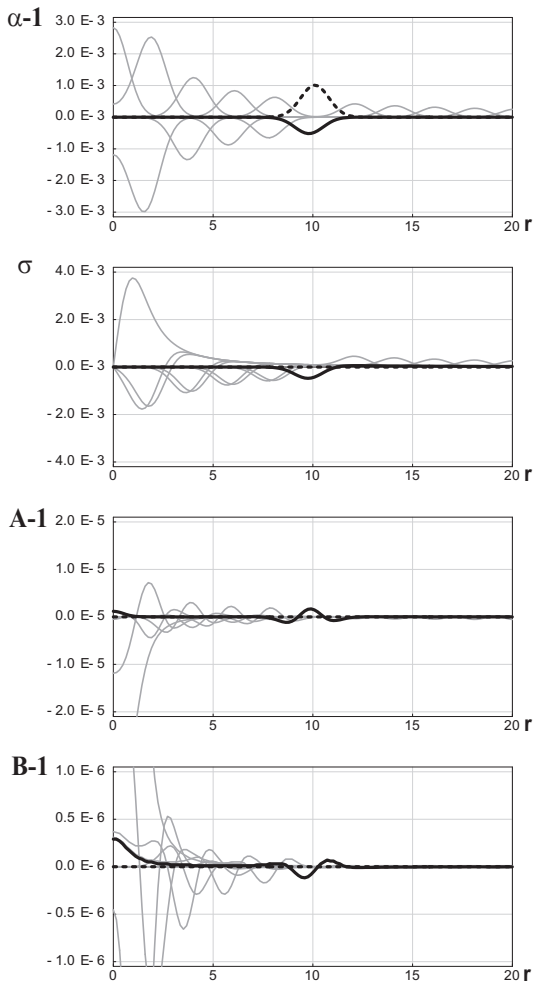}
\caption{As previous plot, but using the harmonic shift condition with
$h=1$.}
\label{fig:sphsymmh1}
\end{figure}

Next we consider the same situation, but now using a harmonic shift
with $h=1$.  Fig.~\ref{fig:sphsymmh1} shows results from this run.
The lapse behaves in exactly the same way as before, but now there is
a non-trivial shift.  The evolution of $\sigma$ indicates that the
shift behaves much in the same way as the lapse, with two pulses
traveling in opposite directions, with the inward moving pulse going
through the origin and then moving out as expected.  The evolution of
the metric variables $A$ and $B$ shows that after the ingoing pulse
goes through the origin and starts moving out, the perturbations on
the metric become very small.  The shift then seems to be having a
similar effect to the one it had in the 1+1 case, making the metric
components evolve toward trivial values.

\begin{figure}
\epsfxsize=70mm
\epsfbox{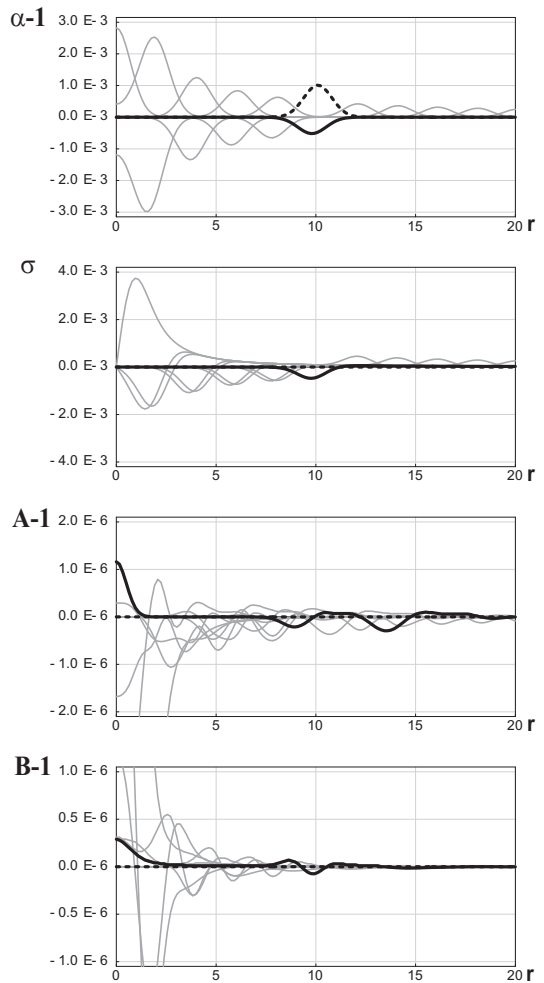}
\caption{Same as previous plots, but now using the generalized
harmonic shift with $h=2$. The metric coefficient $A$ (and to a lesser
extent $B$) shows evidence of two pulses separating and traveling
at different speeds.}
\label{fig:sphsymmh2}
\end{figure}

Fig.~\ref{fig:sphsymmh2} shows a similar run, but now using $f=1$ and
$h=2$.  The whole simulation behaves much the same way as before,
except for the fact that the metric coefficient $A$ (and to a lesser
extent $B$) now shows evidence of two pulses separating and traveling
at different speeds after the rebound through the origin.

\begin{figure}
\epsfxsize=70mm
\epsfbox{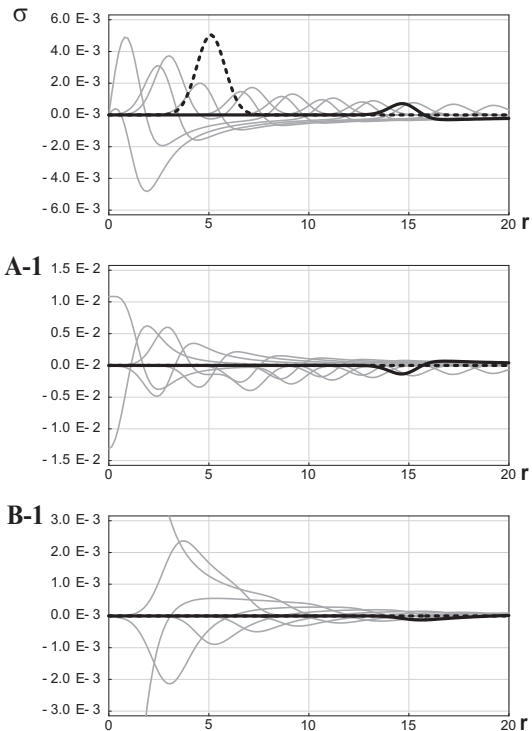}
\caption{Trivial Minkowski slices where the lapse remains equal to unity, 
together with the harmonic shift condition ($h=1$).}
\label{fig:sphsymmnolapse}
\end{figure}

Finally, in Fig.~\ref{fig:sphsymmnolapse} we show a simulation with
$h=1$ for a case where we have left the lapse equal to one throughout the
evolution.  The initial data in this case is purely Minkowski data with a 
shift of the form
\begin{equation}
\sigma = \kappa r \left( e^{- (r - r_c)^2/s^2}
+ e^{- (r + r_c)^2/s^2} \right) \; ,
\end{equation}
with $\kappa = 10^{-3}$, $s = 1$ and $r_c = 5$.  The purpose of this
run is to decouple the harmonic shift condition from the slicing
condition.  The figure shows clearly how, even though the initial
pulse in the shift produces perturbations in the metric coefficients,
these perturbations rapidly decrease in size leaving trivial values
behind them.

One should mention the fact that, even though we don't show
convergence plots in this section, in all cases convergence has been
studied and we have found that the simulations converge at close to
second order.

%%%%%%%%%%%%%%%%%%%%%%
%%%   DISCUSSION   %%%
%%%%%%%%%%%%%%%%%%%%%%

\section{Discussion}
\label{sec:discussion}

We have proposed a natural generalization of the condition for
harmonic spatial coordinates analogous to the generalization of
harmonic time slices of Bona {\em et al.}~\cite{Bona94b}, and closely
related to shift conditions recently introduced by Lindblom and
Scheel~\cite{Lindblom:2003ad}, and by Bona and
Palenzuela~\cite{Bona:2004yp}.  This coordinate condition implies an 
evolution equation for the shift components.  We have also found that 
if one wants to decouple this evolution equation for the shift from 
the choice of slicing condition, it is important to work with a rescaled 
shift vector $\sigma^i = \beta^i / \alpha$.

The generalized harmonic shift condition thus obtained turns out not
to be 3-covariant, which is not surprising as it involves the
3-Christoffel symbols directly.  In order to be able to use this
condition in arbitrary sets of curvilinear coordinates, and to be sure
that we always obtain the same shift independently of the choice of
spatial coordinates, we have proposed that the condition should be
interpreted as always being applied to topologically Cartesian
coordinates, and later rewritten in a general curvilinear coordinate
system.  In this way we have obtained a fully 3-covariant version of
the generalized harmonic shift condition.

We have shown that the evolution equation for the shift proposed here
can be seen to lead to strongly hyperbolic evolution systems both in
the case of 1+1 ``toy'' relativity and in the case of spherical
symmetry.  Though we have not done a completely general analysis here,
it is to be expected that it will also lead to strongly hyperbolic
systems in the 3D case.  Here we have concentrated on simple
one-dimensional systems in order to take the hyperbolicity analysis
further and study the possible formation of blow-ups associated with
this shift condition.  We find that the coefficient $h$ controlling
the gauge speed associated with the shift can be an arbitrary function
of the lapse, but must be independent of the shift itself in order to
avoid blow-ups.  In the slicing and constraint sectors we recover
previous results found in~\cite{Reimann:2004wp}.  An important result
of this study is the fact that evolutions will be much better behaved
if the gauge speeds associated with the lapse and shift are the same.
This can be understood from the fact that terms in the sources that
are mixed products of eigenfields associated to the lapse and shift
vanish in this case.  This implies that if one wants to use a shift of
the generalized harmonic family, together with a lapse of the
Bona-Masso type (like 1+log slicing), it is best to take
$h(\alpha)=f(\alpha)$.  In particular, the shock avoiding family $h=f=
1 + \mbox{const}/\alpha^2$ is an optimal choice.

We have also performed a series of numerical simulation both to
confirm the predictions of the blow-up analysis, and to study what
effect the shift has on the evolution of the geometric variables.  In
the 1+1 dimensional case, we find that the effect of the shift is to
take the spatial metric back to a trivial value everywhere, by
propagating away any non-trivial values in a wavelike fashion.  In
spherical symmetry the situation is considerably richer, but our main
result is that when one uses the 3-covariant version of the
generalized harmonic shift, then the effect of the shift is also to
drive the metric coefficients to trivial values by propagating away
any initial perturbations in the way one would expect for spherical
waves, {\em i.e.} the perturbations become smaller as they propagate
outwards. It is important to mention that, had we not used the
covariant form of the shift condition and tried to apply the original
non-covariant version directly to spherical coordinates, we would have
found the shift condition to be singular, and worse still, to break
the original spherical symmetry of the system.  This shows that
working with the 3-covariant version is the correct approach.

As a final comment, one should also mention the fact that the
requirement of 3-covariance is not satisfied by some recently proposed
shift conditions that are currently being used by large scale 3D
simulations, such as the ``Gamma driver''
shift~\cite{Alcubierre02a,Bruegmann03,Alcubierre:2004hr}.  We are
currently also studying 3-covariant versions of those conditions.

%%%%%%%%%%%%%%%%%%%%%%%%%%%
%%%   ACKNOWLEDGMENTS   %%%
%%%%%%%%%%%%%%%%%%%%%%%%%%%

\acknowledgments

It is a pleasure to thank Sascha Husa for many useful discussions and
comments.  This work was supported in part by DGAPA-UNAM through
grants IN112401, IN122002 and IN119005 and by DFG grant ``SFB Transregio 7:
Gravitationswellenastronomie''.  B.R. acknowledges financial support
from the Richard-Schieber Stiftung. All runs were made at the
\texttt{ekbek} linux cluster in the ``Laboratorio de Super-C\'omputo
Astrof\'{\i}sico (LaSumA)'', CINVESTAV, built through CONACyT grant
42748.

%%%%%%%%%%%%%%%%%%%%%%%%%%%%
%%%   START APPENDICES   %%%
%%%%%%%%%%%%%%%%%%%%%%%%%%%%

\appendix

%%%%%%%%%%%%%%%%%%%%%%%%%%%%%%%%%%%%%%%%%%%%%%
%%%   APPENDIX: MODIFIED LAPSE AND SHIFT   %%%
%%%%%%%%%%%%%%%%%%%%%%%%%%%%%%%%%%%%%%%%%%%%%%

\section{Generalized harmonic lapse and shift conditions}
\label{sec:appendix_modified}

Here we will provide a general derivation of
equations~(\ref{eq:alphaBM}) and~(\ref{eq:betadotGH}) for the lapse
and shift.  Let us start by considering the d'Alambertian of any
number $a$ of functions $\psi^a(x^\mu)$ with their corresponding source
terms
\begin{equation}
\Box \psi^a = S^{\psi^a} \; .
\end{equation}
Now, the d'Alambertian can be written in general as
\begin{equation}
\Box \psi^a = \frac{1}{\sqrt{-g}} \; \partial_\mu
\left[ \sqrt{-g} g^{\mu\nu} \partial_\nu \psi^a \right] \; .
\end{equation}
Using $g^{\mu\nu}= \gamma^{\mu\nu} - n^\mu n^\nu$, with
$\gamma^{\mu\nu}$ the projector operator on the hypersurfaces
$\Sigma_t$ with normal $n^\mu$, we find
\begin{eqnarray}
\Box \psi^a &=& \frac{1}{\alpha \sqrt{\gamma}} \: \partial_\mu \left[
\alpha \sqrt{\gamma} \: \gamma^{\mu\nu} \partial_\nu \psi^a \right]
\nonumber \\
&-& \frac{1}{\alpha \sqrt{\gamma}} \: \partial_\mu \left[
\alpha \sqrt{\gamma} \: n^\mu n^\nu \partial_\nu \psi^a \right] \; ,
\end{eqnarray}
where we used the fact that $g:={\rm det}\:g_{\mu\nu}= - \alpha^2
\gamma$ with $\gamma:={\rm det}\:\gamma_{ij}$ the determinant of the
3-metric on $\Sigma_t$.  We then have
\begin{eqnarray}
\Box \psi^a &=& \,^3\Delta \psi^a + a^\mu \nabla_\mu \psi^a
+ K \: n^\mu\nabla _\mu \psi^a \nonumber \\
&-& n^\mu \nabla _\mu \left( n^\nu \nabla _\nu \psi^a\right) \; ,
\end{eqnarray}
where $^3\Delta$ is the Laplacian compatible with the 3-metric
$\gamma_{ij}$, $a^\mu= n^\nu \nabla_\nu n^\mu\equiv
\gamma^{\mu\nu}\nabla_\nu [{\rm ln} \: \alpha]=: D^\mu [{\rm ln} \:
\alpha]$ is the \mbox{4-acceleration} of the normal observers, and we
used \mbox{$K = - \nabla_\nu n^\nu$}.

In order to obtain for instance a system of first order equations one
can further define
\begin{eqnarray}
\label{Q}
Q^{\mu a}&:=& D^\mu \psi^a \; , \\
\label{Pi}
\Pi^a &:=& {\cal L}_{\vec{n}} \: \psi^a = n^\nu \nabla _\nu \psi^a \; ,
\end{eqnarray}
where $D^\mu \psi^a:= \gamma^{\mu\sigma}\nabla_\sigma
\psi^a$. Collecting the above results we obtain
\begin{equation}
\label{3+1Box}
{\cal L}_{\vec{n}} \: \Pi^a - a_\mu Q^{\mu a} - D_\mu Q^{\mu a}- \Pi^a 
K= -  S^{\psi^a} \; .
\end{equation}

A simple application of the above results is the case when 
$\psi^a = x^a$, in which case $\Pi^a = n^a = (1,-\beta^i)/\alpha$, 
$Q^a_\mu = \gamma^a_\mu$, and $D_\nu Q^{\nu i} = 
\partial_j\left(\sqrt{\gamma} \: \gamma^{ij}\right) /
\sqrt{\gamma} \equiv - {}^{(3)}\Gamma^i$.  The above equation with 
$S^{\psi^a} = 0$ is then called the {\it harmonic coordinate
condition}, which provides an evolution equation for the lapse
- Eq.~(\ref{eq:alphadothar}) - and for the shift - Eq.~(\ref{eq:betadothar}) -
when taking $x^a = (t,x^i)$ with $t$ defining the time slicings and
$x^i$ being spatial coordinates on $\Sigma_t$.  

On the other hand one can take $\Box \psi^a=S^{\psi^a}$ with a source term 
of the form 
\mbox{$S^{\psi^a}= q_{\psi^a} n^\mu n^\nu \nabla_\mu \nabla_\nu \psi^a$} 
(no sum over index $a$). Now, using 
$n^\mu n^\nu= \gamma^{\mu\nu} - g^{\mu\nu}$,
one obtains
\begin{equation}
n^\mu n^\nu \nabla_\mu \nabla_\nu \psi^a = 
\gamma^{\mu\nu} \nabla_\mu \nabla_\nu \psi^a - \Box \psi^a \; .
\end{equation}
Using the orthogonal decomposition 
$\nabla_\nu \psi^a= D_\nu \psi^a - n_\nu n_\sigma \nabla^\sigma \psi^a
= Q_\nu^a - n_\nu \Pi^a$ we find
\begin{equation}
n^\mu n^\nu \nabla_\mu\nabla_\nu \psi^a = D_\nu Q^{\nu a}
+ \Pi^a K - \Box \psi^a \; ,
\end{equation}
where we used $\gamma^{\mu\sigma}\nabla_\mu n_\sigma= \nabla_\mu
n^\mu= -K$ and $\gamma^{\mu\nu} n_\nu\equiv 0$.  In this way the
equation $\Box \psi^a = S^{\psi^a}$ becomes
\begin{equation}
\Box \psi^a = \frac{q_{\psi^a}}{1+q_{\psi^a}}\left(D_\nu Q^{\nu a}
+ \Pi^a K \right) \; .
\end{equation}

Finally, $-\Box \psi^a$ is given by the left hand side of Eq.~(\ref{3+1Box}),
from where we find
\begin{eqnarray}
{\cal L}_{\vec{n}} \Pi^a - a_\mu Q^{\mu a} - D_\mu Q^{\mu a}- \Pi^a K = 
\hspace{10mm} \nonumber \\
- \frac{q_{\psi^a}}{1+q_{\psi^a}}\left(D_\nu Q^{\nu a} + \Pi^a K \right) \; ,
\end{eqnarray}
which simplifies to
\begin{equation}
\label{3+1BoxS}
{\cal L}_{\vec{n}} \Pi^a - a_\mu Q^{\mu a}
= \frac{1}{1+q_{\psi^a}}\left(D_\nu Q^{\nu a} + \Pi^a K \right) \; .
\end{equation}

In this way by taking $\psi^a=(t,x^i)$, $q_t=a_f=1/f-1$,
$q_{x^i}=a_h=1/h-1$, together with Eqs.~(\ref{Q}) and~(\ref{Pi})
(leading to $\Pi^a=n^a=(1,-\beta^i)/\alpha$ and
\mbox{$Q^a_\mu=\gamma_\mu^a$}), one recovers the evolution
equations~(\ref{eq:alphaBM}) and~(\ref{eq:betadotGH}) for $\alpha$ and
$\beta^i$, respectively.

%%%%%%%%%%%%%%%%%%%%%%%%%%%%%%%%%%%%%%%%%%%%
%%%   APPENDIX: METRIC AND CHRISTOFFEL   %%%
%%%%%%%%%%%%%%%%%%%%%%%%%%%%%%%%%%%%%%%%%%%%

\section{Metric and Christoffel symbols in 3+1 language}
\label{sec:appendix_Christoffel}

For the expression of the generalized harmonic gauge conditions one
needs to write the 4-metric of spacetime and its associated
Christoffel symbols in 3+1 language.  The 4-metric in terms of 3+1
quantities has the form
\begin{eqnarray}
g_{00} &=& -\left( \alpha^2 - \gamma_{ij}\,\beta^i \beta^j \right) \; , \\
g_{0i} &=& \gamma_{ij}\,\beta^j \; , \\
g_{ij} &=& \gamma_{ij} \; ,
\end{eqnarray}
and the corresponding inverse metric is
\begin{eqnarray}
g^{00} &=& - 1 / \alpha^2 \; , \\
g^{0i} &=& \beta^i / \alpha^2 \; , \\
g^{ij} &=& \gamma^{ij} - \beta^i \beta^j / \alpha^2 \; .
\end{eqnarray}

From this one can obtain the following expressions for the
4-Christoffel symbols in terms of 3+1 quantities
\begin{eqnarray}
\Gamma_{00}^0 &=& \left( \partial_t\,\alpha +
\beta^m \partial_m\,\alpha - \beta^m \beta^n K_{mn} \right) / \alpha \; , \\
\Gamma_{0i}^0 &=& \left( \partial_i \alpha -
\beta^m K_{im} \right) / \alpha \; , \\
\Gamma_{ij}^0 &=& - K_{ij} / \alpha \; , \\
\Gamma_{00}^l &=& \alpha \partial^l \alpha - 2 \alpha \beta^m K_m^l
\nonumber \\
&-& \beta^l \left( \partial_t \alpha +
\beta^m \partial_m \alpha - \beta^m \beta^n K_{mn} \right) / 
\alpha \nonumber \\
&+& \partial_t \beta^l + \beta^m \; {}^{(3)}\nabla_m \beta^l \; , \\
\Gamma_{m0}^l &=& - \beta^l \left( \partial_m \alpha -
\beta^n K_{mn} \right) / \alpha \nonumber \\
&-& \alpha K^l_m + {}^{(3)}\nabla_m \beta^l \; , \\
\Gamma_{ij}^l &=& {}^{(3)}\Gamma_{ij}^l + \beta^l K_{ij} / \alpha \; .
\end{eqnarray}

\pagebreak

The contracted Christoffel symbols $\Gamma^\alpha := g^{\mu \nu}
\Gamma^\alpha_{\mu \nu}$ then become
\begin{equation}
\Gamma^0 = - \frac{1}{\alpha^3}
\left( \partial_t \alpha - \beta^m \partial_m \alpha + \alpha^2 K \right) 
\end{equation}

\pagebreak

and
\begin{eqnarray}
\Gamma^i &=& \frac{\beta^i}{\alpha^3} \left( \partial_t \alpha
- \beta^m \partial_m \alpha + \alpha^2 K \right) 
+ {}^{(3)}\Gamma^i \nonumber \\
&-& \frac{1}{\alpha^2} \left( \partial_t \beta^i - \beta^m \partial_m \beta^i
+ \alpha \partial^i \alpha  \right) \; .
\end{eqnarray}

%%%%%%%%%%%%%%%%%%%%%%
%%%   REFERENCES   %%%
%%%%%%%%%%%%%%%%%%%%%%

\bibliographystyle{bibtex/apsrev}
\bibliography{bibtex/referencias}

\begin{thebibliography}{33}
\expandafter\ifx\csname natexlab\endcsname\relax\def\natexlab#1{#1}\fi
\expandafter\ifx\csname bibnamefont\endcsname\relax
  \def\bibnamefont#1{#1}\fi
\expandafter\ifx\csname bibfnamefont\endcsname\relax
  \def\bibfnamefont#1{#1}\fi
\expandafter\ifx\csname citenamefont\endcsname\relax
  \def\citenamefont#1{#1}\fi
\expandafter\ifx\csname url\endcsname\relax
  \def\url#1{\texttt{#1}}\fi
\expandafter\ifx\csname urlprefix\endcsname\relax\def\urlprefix{URL }\fi
\providecommand{\bibinfo}[2]{#2}
\providecommand{\eprint}[2][]{\url{#2}}

\bibitem[{\citenamefont{Smarr and York}(1978{\natexlab{a}})}]{Smarr78a}
\bibinfo{author}{\bibfnamefont{L.}~\bibnamefont{Smarr}} \bibnamefont{and}
  \bibinfo{author}{\bibfnamefont{J.}~\bibnamefont{York}},
  \bibinfo{journal}{Phys. Rev. D} \textbf{\bibinfo{volume}{17}},
  \bibinfo{pages}{1945} (\bibinfo{year}{1978}{\natexlab{a}}).

\bibitem[{\citenamefont{Smarr and York}(1978{\natexlab{b}})}]{Smarr78b}
\bibinfo{author}{\bibfnamefont{L.}~\bibnamefont{Smarr}} \bibnamefont{and}
  \bibinfo{author}{\bibfnamefont{J.}~\bibnamefont{York}},
  \bibinfo{journal}{Phys. Rev. D} \textbf{\bibinfo{volume}{17}},
  \bibinfo{pages}{2529} (\bibinfo{year}{1978}{\natexlab{b}}).

\bibitem[{\citenamefont{Sarbach and Tiglio}(2002)}]{Sarbach:2002gr}
\bibinfo{author}{\bibfnamefont{O.}~\bibnamefont{Sarbach}} \bibnamefont{and}
  \bibinfo{author}{\bibfnamefont{M.}~\bibnamefont{Tiglio}},
  \bibinfo{journal}{Phys. Rev.} \textbf{\bibinfo{volume}{D66}},
  \bibinfo{pages}{064023} (\bibinfo{year}{2002}), \eprint{gr-qc/0205086}.

\bibitem[{\citenamefont{Alcubierre et~al.}(2002)\citenamefont{Alcubierre,
  Br\"ugmann, Diener, Koppitz, Pollney, Seidel, and Takahashi}}]{Alcubierre02a}
\bibinfo{author}{\bibfnamefont{M.}~\bibnamefont{Alcubierre}},
  \bibinfo{author}{\bibfnamefont{B.}~\bibnamefont{Br\"ugmann}},
  \bibinfo{author}{\bibfnamefont{P.}~\bibnamefont{Diener}},
  \bibinfo{author}{\bibfnamefont{M.}~\bibnamefont{Koppitz}},
  \bibinfo{author}{\bibfnamefont{D.}~\bibnamefont{Pollney}},
  \bibinfo{author}{\bibfnamefont{E.}~\bibnamefont{Seidel}}, \bibnamefont{and}
  \bibinfo{author}{\bibfnamefont{R.}~\bibnamefont{Takahashi}},
  \bibinfo{journal}{Phys. Rev. D} \textbf{\bibinfo{volume}{67}},
  \bibinfo{pages}{084023} (\bibinfo{year}{2002}),
  \bibinfo{note}{gr-qc/0206072}.

\bibitem[{\citenamefont{Alcubierre}(2003)}]{Alcubierre02b}
\bibinfo{author}{\bibfnamefont{M.}~\bibnamefont{Alcubierre}},
  \bibinfo{journal}{Class. Quantum Grav.} \textbf{\bibinfo{volume}{20}},
  \bibinfo{pages}{607} (\bibinfo{year}{2003}), \bibinfo{note}{gr-qc/0210050}.

\bibitem[{\citenamefont{Alcubierre
  et~al.}(2003{\natexlab{a}})\citenamefont{Alcubierre, Corichi, Gonz\'alez,
  Nu{\~n}ez, and Salgado}}]{Alcubierre03b}
\bibinfo{author}{\bibfnamefont{M.}~\bibnamefont{Alcubierre}},
  \bibinfo{author}{\bibfnamefont{A.}~\bibnamefont{Corichi}},
  \bibinfo{author}{\bibfnamefont{J.}~\bibnamefont{Gonz\'alez}},
  \bibinfo{author}{\bibfnamefont{D.}~\bibnamefont{Nu{\~n}ez}},
  \bibnamefont{and} \bibinfo{author}{\bibfnamefont{M.}~\bibnamefont{Salgado}},
  \bibinfo{journal}{Class. Quant. Grav} \textbf{\bibinfo{volume}{20}},
  \bibinfo{pages}{3951} (\bibinfo{year}{2003}{\natexlab{a}}),
  \bibinfo{note}{gr-qc/0303069}.

\bibitem[{\citenamefont{Alcubierre
  et~al.}(2003{\natexlab{b}})\citenamefont{Alcubierre, Corichi, Gonz\'alez,
  Nu{\~n}ez, and Salgado}}]{Alcubierre03c}
\bibinfo{author}{\bibfnamefont{M.}~\bibnamefont{Alcubierre}},
  \bibinfo{author}{\bibfnamefont{A.}~\bibnamefont{Corichi}},
  \bibinfo{author}{\bibfnamefont{J.}~\bibnamefont{Gonz\'alez}},
  \bibinfo{author}{\bibfnamefont{D.}~\bibnamefont{Nu{\~n}ez}},
  \bibnamefont{and} \bibinfo{author}{\bibfnamefont{M.}~\bibnamefont{Salgado}},
  \bibinfo{journal}{Phys. Rev. D} \textbf{\bibinfo{volume}{67}},
  \bibinfo{pages}{104021} (\bibinfo{year}{2003}{\natexlab{b}}),
  \bibinfo{note}{gr-qc/0303086}.

\bibitem[{\citenamefont{Lindblom and Scheel}(2003)}]{Lindblom:2003ad}
\bibinfo{author}{\bibfnamefont{L.}~\bibnamefont{Lindblom}} \bibnamefont{and}
  \bibinfo{author}{\bibfnamefont{M.~A.} \bibnamefont{Scheel}},
  \bibinfo{journal}{Phys. Rev.} \textbf{\bibinfo{volume}{D67}},
  \bibinfo{pages}{124005} (\bibinfo{year}{2003}), \eprint{gr-qc/0301120}.

\bibitem[{\citenamefont{Bruhat}(1952)}]{Choquet52}
\bibinfo{author}{\bibfnamefont{Y.}~\bibnamefont{Bruhat}},
  \bibinfo{journal}{Acta Mathematica} \textbf{\bibinfo{volume}{88}},
  \bibinfo{pages}{141} (\bibinfo{year}{1952}).

\bibitem[{\citenamefont{Szilagyi et~al.}(2002)\citenamefont{Szilagyi, Schmidt,
  and Winicour}}]{Szilagyi02b}
\bibinfo{author}{\bibfnamefont{B.}~\bibnamefont{Szilagyi}},
  \bibinfo{author}{\bibfnamefont{B.}~\bibnamefont{Schmidt}}, \bibnamefont{and}
  \bibinfo{author}{\bibfnamefont{J.}~\bibnamefont{Winicour}},
  \bibinfo{journal}{Phys. Rev. D} \textbf{\bibinfo{volume}{65}}
  (\bibinfo{year}{2002}), \bibinfo{note}{gr-qc/0106026}.

\bibitem[{\citenamefont{Garfinkle}(2002)}]{Garfinkle02a}
\bibinfo{author}{\bibfnamefont{D.}~\bibnamefont{Garfinkle}},
  \bibinfo{journal}{Phys. Rev. D} \textbf{\bibinfo{volume}{65}},
  \bibinfo{pages}{044029} (\bibinfo{year}{2002}),
  \bibinfo{note}{gr-qc/0110013}.

\bibitem[{\citenamefont{Pretorius}(2005)}]{Pretorius:2004jg}
\bibinfo{author}{\bibfnamefont{F.}~\bibnamefont{Pretorius}},
  \bibinfo{journal}{Class. Quant. Grav.} \textbf{\bibinfo{volume}{22}},
  \bibinfo{pages}{425} (\bibinfo{year}{2005}), \eprint{gr-qc/0407110}.

\bibitem[{\citenamefont{Bona and Mass\'{o}}(1988)}]{Bona88}
\bibinfo{author}{\bibfnamefont{C.}~\bibnamefont{Bona}} \bibnamefont{and}
  \bibinfo{author}{\bibfnamefont{J.}~\bibnamefont{Mass\'{o}}},
  \bibinfo{journal}{Phys. Rev. D} \textbf{\bibinfo{volume}{38}},
  \bibinfo{pages}{2419} (\bibinfo{year}{1988}).

\bibitem[{\citenamefont{Bona and Mass\'{o}}(1989)}]{Bona89}
\bibinfo{author}{\bibfnamefont{C.}~\bibnamefont{Bona}} \bibnamefont{and}
  \bibinfo{author}{\bibfnamefont{J.}~\bibnamefont{Mass\'{o}}},
  \bibinfo{journal}{Phys. Rev. D} \textbf{\bibinfo{volume}{40}},
  \bibinfo{pages}{1022} (\bibinfo{year}{1989}).

\bibitem[{\citenamefont{Bona and Mass\'{o}}(1992)}]{Bona92}
\bibinfo{author}{\bibfnamefont{C.}~\bibnamefont{Bona}} \bibnamefont{and}
  \bibinfo{author}{\bibfnamefont{J.}~\bibnamefont{Mass\'{o}}},
  \bibinfo{journal}{Phys. Rev. Lett.} \textbf{\bibinfo{volume}{68}},
  \bibinfo{pages}{1097} (\bibinfo{year}{1992}).

\bibitem[{\citenamefont{Abrahams et~al.}(1995)\citenamefont{Abrahams, Anderson,
  Choquet-Bruhat, and York}}]{Abrahams95a}
\bibinfo{author}{\bibfnamefont{A.}~\bibnamefont{Abrahams}},
  \bibinfo{author}{\bibfnamefont{A.}~\bibnamefont{Anderson}},
  \bibinfo{author}{\bibfnamefont{Y.}~\bibnamefont{Choquet-Bruhat}},
  \bibnamefont{and} \bibinfo{author}{\bibfnamefont{J.}~\bibnamefont{York}},
  \bibinfo{journal}{Phys. Rev. Lett.} \textbf{\bibinfo{volume}{75}},
  \bibinfo{pages}{3377} (\bibinfo{year}{1995}), \bibinfo{note}{gr-qc/9506072}.

\bibitem[{\citenamefont{Geyer and Herold}(1995)}]{Geyer95}
\bibinfo{author}{\bibfnamefont{A.}~\bibnamefont{Geyer}} \bibnamefont{and}
  \bibinfo{author}{\bibfnamefont{H.}~\bibnamefont{Herold}},
  \bibinfo{journal}{Phys. Rev. D} \textbf{\bibinfo{volume}{52}},
  \bibinfo{pages}{6182} (\bibinfo{year}{1995}).

\bibitem[{\citenamefont{Bona et~al.}(1997)\citenamefont{Bona, Mass{\'o},
  Seidel, and Stela}}]{Bona97a}
\bibinfo{author}{\bibfnamefont{C.}~\bibnamefont{Bona}},
  \bibinfo{author}{\bibfnamefont{J.}~\bibnamefont{Mass{\'o}}},
  \bibinfo{author}{\bibfnamefont{E.}~\bibnamefont{Seidel}}, \bibnamefont{and}
  \bibinfo{author}{\bibfnamefont{J.}~\bibnamefont{Stela}},
  \bibinfo{journal}{Phys. Rev. D} \textbf{\bibinfo{volume}{56}},
  \bibinfo{pages}{3405} (\bibinfo{year}{1997}), \bibinfo{note}{gr-qc/9709016}.

\bibitem[{\citenamefont{Bona et~al.}(1995)\citenamefont{Bona, Mass{\'o},
  Seidel, and Stela}}]{Bona94b}
\bibinfo{author}{\bibfnamefont{C.}~\bibnamefont{Bona}},
  \bibinfo{author}{\bibfnamefont{J.}~\bibnamefont{Mass{\'o}}},
  \bibinfo{author}{\bibfnamefont{E.}~\bibnamefont{Seidel}}, \bibnamefont{and}
  \bibinfo{author}{\bibfnamefont{J.}~\bibnamefont{Stela}},
  \bibinfo{journal}{Phys. Rev. Lett.} \textbf{\bibinfo{volume}{75}},
  \bibinfo{pages}{600} (\bibinfo{year}{1995}), \bibinfo{note}{gr-qc/9412071}.

\bibitem[{\citenamefont{Bernstein}(1993)}]{Bernstein93a}
\bibinfo{author}{\bibfnamefont{D.}~\bibnamefont{Bernstein}}, Ph.D. thesis,
  \bibinfo{school}{University of Illinois Urbana-Champaign}
  (\bibinfo{year}{1993}).

\bibitem[{\citenamefont{Anninos et~al.}(1995)\citenamefont{Anninos, Camarda,
  Mass{\'o}, Seidel, Suen, and Towns}}]{Anninos94c}
\bibinfo{author}{\bibfnamefont{P.}~\bibnamefont{Anninos}},
  \bibinfo{author}{\bibfnamefont{K.}~\bibnamefont{Camarda}},
  \bibinfo{author}{\bibfnamefont{J.}~\bibnamefont{Mass{\'o}}},
  \bibinfo{author}{\bibfnamefont{E.}~\bibnamefont{Seidel}},
  \bibinfo{author}{\bibfnamefont{W.-M.} \bibnamefont{Suen}}, \bibnamefont{and}
  \bibinfo{author}{\bibfnamefont{J.}~\bibnamefont{Towns}},
  \bibinfo{journal}{Phys. Rev. D} \textbf{\bibinfo{volume}{52}},
  \bibinfo{pages}{2059} (\bibinfo{year}{1995}).

\bibitem[{\citenamefont{Bona and Palenzuela}(2004)}]{Bona:2004yp}
\bibinfo{author}{\bibfnamefont{C.}~\bibnamefont{Bona}} \bibnamefont{and}
  \bibinfo{author}{\bibfnamefont{C.}~\bibnamefont{Palenzuela}},
  \bibinfo{journal}{Phys. Rev.} \textbf{\bibinfo{volume}{D69}},
  \bibinfo{pages}{104003} (\bibinfo{year}{2004}), \eprint{gr-qc/0401019}.

\bibitem[{\citenamefont{York}(1979)}]{York79}
\bibinfo{author}{\bibfnamefont{J.}~\bibnamefont{York}}, in
  \emph{\bibinfo{booktitle}{Sources of Gravitational Radiation}}, edited by
  \bibinfo{editor}{\bibfnamefont{L.}~\bibnamefont{Smarr}}
  (\bibinfo{publisher}{Cambridge University Press},
  \bibinfo{address}{Cambridge, England}, \bibinfo{year}{1979}).

\bibitem[{\citenamefont{Bona and Mass{\'o}}(1993)}]{Bona93}
\bibinfo{author}{\bibfnamefont{C.}~\bibnamefont{Bona}} \bibnamefont{and}
  \bibinfo{author}{\bibfnamefont{J.}~\bibnamefont{Mass{\'o}}},
  \bibinfo{journal}{International Journal of Modern Physics C: Physics and
  Computers} \textbf{\bibinfo{volume}{4}}, \bibinfo{pages}{88}
  (\bibinfo{year}{1993}).

\bibitem[{\citenamefont{Alcubierre}(1997)}]{Alcubierre97a}
\bibinfo{author}{\bibfnamefont{M.}~\bibnamefont{Alcubierre}},
  \bibinfo{journal}{Phys. Rev. D} \textbf{\bibinfo{volume}{55}},
  \bibinfo{pages}{5981} (\bibinfo{year}{1997}), \bibinfo{note}{gr-qc/9609015}.

\bibitem[{\citenamefont{Reimann et~al.}(2005)\citenamefont{Reimann, Alcubierre,
  Gonz\'alez, and Nu{\~n}ez}}]{Reimann:2004wp}
\bibinfo{author}{\bibfnamefont{B.}~\bibnamefont{Reimann}},
  \bibinfo{author}{\bibfnamefont{M.}~\bibnamefont{Alcubierre}},
  \bibinfo{author}{\bibfnamefont{J.~A.} \bibnamefont{Gonz\'alez}},
  \bibnamefont{and}
  \bibinfo{author}{\bibfnamefont{D.}~\bibnamefont{Nu{\~n}ez}},
  \bibinfo{journal}{Phys. Rev.} \textbf{\bibinfo{volume}{D71}},
  \bibinfo{pages}{064021} (\bibinfo{year}{2005}), \eprint{gr-qc/0411094}.

\bibitem[{\citenamefont{Alinhac}(1995)}]{Alinhac}
\bibinfo{author}{\bibfnamefont{S.}~\bibnamefont{Alinhac}},
  \emph{\bibinfo{title}{Blowup for Nonlinear Hyperbolic Equations}}
  (\bibinfo{publisher}{Birkh\"auser}, \bibinfo{address}{Boston., U.S.A.},
  \bibinfo{year}{1995}).

\bibitem[{\citenamefont{John}(1986)}]{John86}
\bibinfo{author}{\bibfnamefont{F.}~\bibnamefont{John}},
  \emph{\bibinfo{title}{Partial Differential Equations}}
  (\bibinfo{publisher}{Springer-Verlag}, \bibinfo{address}{New York},
  \bibinfo{year}{1986}).

\bibitem[{\citenamefont{Arnowitt et~al.}(1962)\citenamefont{Arnowitt, Deser,
  and Misner}}]{Arnowitt62}
\bibinfo{author}{\bibfnamefont{R.}~\bibnamefont{Arnowitt}},
  \bibinfo{author}{\bibfnamefont{S.}~\bibnamefont{Deser}}, \bibnamefont{and}
  \bibinfo{author}{\bibfnamefont{C.~W.} \bibnamefont{Misner}}, in
  \emph{\bibinfo{booktitle}{Gravitation: An Introduction to Current Research}},
  edited by \bibinfo{editor}{\bibfnamefont{L.}~\bibnamefont{Witten}}
  (\bibinfo{publisher}{John Wiley}, \bibinfo{address}{New York},
  \bibinfo{year}{1962}), pp. \bibinfo{pages}{227--265}.

\bibitem[{\citenamefont{Alcubierre and Mass{\'o}}(1998)}]{Alcubierre97b}
\bibinfo{author}{\bibfnamefont{M.}~\bibnamefont{Alcubierre}} \bibnamefont{and}
  \bibinfo{author}{\bibfnamefont{J.}~\bibnamefont{Mass{\'o}}},
  \bibinfo{journal}{Phys. Rev. D} \textbf{\bibinfo{volume}{57}},
  \bibinfo{pages}{4511} (\bibinfo{year}{1998}).

\bibitem[{\citenamefont{Alcubierre and Gonz\'alez}(2005)}]{Alcubierre04a}
\bibinfo{author}{\bibfnamefont{M.}~\bibnamefont{Alcubierre}} \bibnamefont{and}
  \bibinfo{author}{\bibfnamefont{J.}~\bibnamefont{Gonz\'alez}},
  \bibinfo{journal}{Comp. Phys. Comm.} \textbf{\bibinfo{volume}{167}},
  \bibinfo{pages}{76} (\bibinfo{year}{2005}), \bibinfo{note}{gr-qc/0401113}.

\bibitem[{\citenamefont{Br{\"u}gmann et~al.}(2004)\citenamefont{Br{\"u}gmann,
  Tichy, and Jansen}}]{Bruegmann03}
\bibinfo{author}{\bibfnamefont{B.}~\bibnamefont{Br{\"u}gmann}},
  \bibinfo{author}{\bibfnamefont{W.}~\bibnamefont{Tichy}}, \bibnamefont{and}
  \bibinfo{author}{\bibfnamefont{N.}~\bibnamefont{Jansen}},
  \bibinfo{journal}{Phys. Rev. Lett.} \textbf{\bibinfo{volume}{92}},
  \bibinfo{pages}{211101} (\bibinfo{year}{2004}),
  \bibinfo{note}{gr-qc/0312112}.

\bibitem[{\citenamefont{Alcubierre et~al.}(2004)}]{Alcubierre:2004hr}
\bibinfo{author}{\bibfnamefont{M.}~\bibnamefont{Alcubierre}}
  \bibnamefont{et~al.} (\bibinfo{year}{2004}), \eprint{gr-qc/0411149}.

\end{thebibliography}

%%%%%%%%%%%%%%%
%%%   END   %%%
%%%%%%%%%%%%%%%

\end{document}